\documentclass[letterpaper]{JHEP3}
\pdfoutput=1

\usepackage{graphicx}
\usepackage{amsmath,amsthm, amssymb}

\newcommand{\eq}{\begin{equation}}
\newcommand{\eqe}{\end{equation}}

\newcommand{\eqa}{\begin{eqnarray}}
\newcommand{\eqae}{\end{eqnarray}}

\title{Dual Superconformal Symmetry of $\mathcal{N}=6$ Chern-Simons Theory. }
\author{Yu-tin Huang\footnote{Email: yhuang@physics.ucla.edu}$^{~1}$, Arthur E. Lipstein \footnote{Email: arthur@theory.caltech.edu}$^{~2}$
\\ \\
\\
\it $^1$ Department of Physics and Astronomy,\\ UCLA,\\
Los Angeles, CA 90095-1547, USA
\\ \\
\\
\it $^2$ California Institute of Technology,\\
Pasadena, CA 91125, USA\;}
\abstract{We demonstrate that the four and six-point tree-level amplitudes of $\mathcal{N}=6$ superconformal Chern-Simons theory (ABJM) enjoy $OSp(6|4)$ dual superconformal symmetry if one enlarges the dual superspace to include three additional Grassmann-even coordinates which correspond to an abelian isometry of CP$^3$. The inclusion of these coordinates enables us to match the nontrivial dual superconformal generators with level-one Yangian generators when acting on on-shell amplitudes. We also discuss some implications of dual conformal symmetry for loop-level amplitudes.}
\preprint{CALT 68-2799, UCLA-TEP-10-105}
\keywords{amplitudes, supertwistor, superconformal}

\begin{document}

\numberwithin{equation}{section}
\section{Introduction}
The scattering amplitudes of $\mathcal{N}=4$ super Yang-Mills theory exhibit many hidden structures~\cite{Bern:2005iz} which are related to type IIB string theory on  $AdS_5 \times S^5$ through the AdS$_5$/CFT$_4$ duality~\cite{AdsCFT}. For example, if one uses the momenta of a given scattering amplitude to define points in a dual space via $p_{i}=x_{i}-x_{i+1}$, then it turns out that the scattering amplitude is related to a light-like polygonal Wilson loop whose cusps are located at the dual points $x_i$. This duality was first proposed at strong coupling as a consequence of the self-T-duality of type IIB string theory on AdS$_5\times$S$^5$~\cite{Alday:2007hr}. Remarkably, this duality also holds at weak coupling~\cite{Drummond:2007aua,Brandhuber:2007yx,Drummond:2007cf}, albeit between light-like polygonal Wilson loops and planar MHV amplitudes. Since the Wilson loop in $\mathcal{N}=4$ super Yang-Mills enjoys conformal symmetry, the duality implies a hidden dual conformal symmetry of the scattering amplitudes which is inequivalent to the original conformal symmetry. Furthermore, once the amplitudes are written in a dual chiral superspace, dual superconformal invariance becomes manifest~\cite{Dualc,restore}.

The presence of dual superconformal symmetry in $\mathcal{N}=4$ sYM is intimately related to its integrability. In particular, by commuting the original and dual superconformal symmetries, one can generate an infinite set of classical symmetries which obey a nonabelian aglebra called the Yangian ~\cite{WittenNappi,Drummond:2009fd}. The original superconformal symmetry generators correspond to the level-0 Yangian generators and the dual superconformal symmetry generators provide part of the level-1 Yangian generators. The infinite set of charges that give rise to Yangian symmetry was first discovered at strong coupling in the sigma model framework \cite{sigma}. One of the most important consequences of Yangian symmetry is that the spectrum of long single-trace operators can be computed to arbitrary order in the 't Hooft coupling in the planar limit using an all-loop Bethe-ansatz \cite{n4int}.

The three-dimensional superconformal Chern-Simons theory recently discovered by Aharony, Bergman, Jafferis, and Maldacena (ABJM) is also believed to be integrable. This theory has $OSp(6|4)$ superconformal symmetry and $U(N)\times U(N)$ gauge symmetry. When $k\ll N\ll k^{5}$ (where $k$ is the Chern-Simons level), it is dual to type IIA string theory on $AdS_{4}\times CP^3$ \cite{Aharony:2008ug}. Classical integrability was demonstrated on the string theory side \cite{Stefanski:2008ik}, the planar dilatation operator in the gauge theory side was shown to be integrable up to six loops ~\cite{Minahan:2008hf}, and an all-loop Bethe ansatz was proposed in \cite{ac}. On the other hand, a discrepancy was found between string theory calculations and the all-loop Bethe ansatz \cite{Krishnan:2008zs} \footnote{Note that several ways to resolve this discrepancy were proposed in \cite{Gromov:2008fy,McLoughlin:2008he,Bandres:2009kw}.} and there are very few results regarding Yangian symmetry of scattering amplitudes. Indeed, it was only until recently that the spinor helicity formalism was developed for three dimensions and applied to various superconformal theories~\cite{Bargheer:2009qu,Bargheer:2010hn,Huang:2010rn}. In particular the authors of \cite{Bargheer:2010hn} demonstrated that the four and six-point tree-level amplitudes of the ABJM theory are Yangian invariant. From our experience with Yangian symmetry in $\mathcal{N}=4$ SYM, it is then natural to ask if the Yangian symmetry can be traced back to a hidden dual superconformal symmetry of the amplitudes.

A related question is whether the ABJM theory exhibits a duality relating scattering amplitudes to null-polygonal Wilson loops, since this would also imply dual superconformal symmetry. The four-cusp null-polygonal Wilson loop was computed to two loops in ~\cite{Henn}, where it was shown that the one-loop contribution vanishes and the two loop contribution has the same form as the one-loop correction to the four-cusp null-polygonal Wilson loop of $\mathcal{N}=4$ sYM. Although the two-loop correction to the four-point ABJM amplitude hasn't been calculated, the one-loop result trivially agrees with that of the four-cusp Wilson loop~\cite{Bargheer:2009qu}.

In $\mathcal{N}=4$ SYM, the Wilson-loop/amplitude duality is a consequence of the fact that type IIB string theory on AdS$_5\times$S$^5$ is self-dual after performing T-dualities along the translational directions of $AdS_{5}$ and fermionic T-duality transformations which restore the Ramond-Ramond and dilaton fields to their original values (without altering the background metric)~\cite{Tdual}. This set of dualities exchanges dual superconformal symmetry with ordinary superconformal symmetry \cite{Drummond:2010qh}. A similar analysis for type IIA string theory in AdS$_4\times$CP$^3$ leads to the conclusion that if one only T-dualizes the three translational directions of $AdS_4$, it is not possible to T-dualize the fermionic sector~\cite{Adam:2009kt,Grassi:2009yj}.

In this note, we will demonstrate that the four-point and six-point tree-level amplitudes of the ABJM theory have $OSp(6|4)$ dual superconformal symmetry. In doing so, we will discover that one has to enlarge the dual space to include three additional Grassmann-even coordinates in order to define the dual generators. The need for three new dual coordinates was first suggested by the analysis of the $OSp(6|4)$ algebra in \cite{Bargheer:2010hn}. Here, we express the dual coordinates in terms of the on-shell variables of the amplitudes and use them to construct the dual superconformal generators. These new coordinates are related to the dual R-symmetry and suggest that type IIA string theory on $AdS_4 \times CP^3$ may be self-dual if one T-dualizes three directions in $CP^3$ in addition to the translational directions of $AdS_4$. By matching the Killing vectors of $CP^3$ with the R-symmetry generators, we find that the directions in $CP^3$ which need to be T-dualized are complex, so it not clear how to T-dualize these directions.

We proceed as follows. In the next section, we review the construction of ABJM amplitudes using the spinor-helicity formalism. In sections 3 and 4 we show that the four-point amplitude satisfies dual conformal symmetry by translating from on-shell space, parameterized by $(\lambda_i^\alpha,\;\eta_i^A)$, to the dual superspace $(x_i^{\alpha\beta},\;\theta_i^{\alpha A})$, where $\alpha=1,2$, $A=1,2,3$, and $i$ labels each external particle. In section 5, we show that the dual conformal boost generator is equivalent to the momentum level-1 generator of the Yangian algebra (when acting on on-shell amplitudes). It follows that the six-point amplitude also has dual conformal symmetry, since it was previously shown to have Yangian symmetry.

In section 6, we attempt to define dual supersymmetry generators, and encounter a problem. The obstacle lies in the fact that half of the dual supersymmetry generators fail to commute with the equations that define the hypersurface in the dual space on which the amplitudes have support:
$$ x_i^{\alpha\beta}-x_{i+1}^{\alpha\beta}=p^{\alpha\beta}_i=\lambda^\alpha_i\lambda^\beta_i$$
$$\theta^{A\alpha}_i-\theta^{A\alpha}_{i+1}=q_i^{A\alpha}=\lambda_i^\alpha\eta_i^A.$$
The dual supersymmetry is ``anomalous" in the sense that one cannot make all of the dual supersymmetry generators consistent with these constraints. We remedy this problem by introducing three new Grassmann-even coordinates: 
$$y^{AB}_i-y^{AB}_{i+1}=r_i^{AB}=\eta_i^A\eta_i^B.$$
These coordinates carry only R-symmetry indices and parameterize the half-coset SU(4)/U(3)$_+$. The lower index $+$ means that we are only considering the coset generators that are positively charged under the U(1) of the isotropy group.\footnote{The U(1) here refers to the one in U(3) which is not part of the SU(3).}

By extending the dual space to include these new coordinates, we are able to construct dual superconformal generators which commute with all of the hyperplane constraints. Furthermore, we show that the dual special supersymmetry and R-symmetry generators are equivalent to level-1 Yangian generators when acting on on-shell amplitudes. Since the remaining dual superconformal generators are trivially related to the ordinary superconformal generators, this implies that the four and six-point tree-level amplitudes of the ABJM theory are invariant under dual conformal symmetry (since they were already shown to be invariant under Yangian symmetry).

In section 7 we analyze the implications of dual conformal invariance on loop amplitudes. Assuming that the planar loop-level amplitudes of ABJM have dual conformal symmetry prior to regularization (which was the case for most amplitudes in $\mathcal{N}=4$ super Yang-Mills ~\cite{Drummond:2006rz, Bern:2006ew, Bern:2007ct, Drummond:2007aua}), we find that the one-loop four-point amplitude must vanish (which is consistent with parity) and we obtain some two loop predictions. In section 8, we present our conclusions.

\section{ABJM amplitudes}
The ABJM theory is a three-dimensional twisted Chern-Simons theory with bi-fundamental matter. The field content consists of four complex scalars $Z^{A}$ and four Dirac fermions $\psi_A$ transforming in the bi-fundamental representation of $U(N)\times U(N)$ (with $A$ running from 1 to 4), as well as two $U(N)$ gauge fields $A_{\mu}$ and $\hat{A}_{\mu}$. The matter fields transform in the fundamental representation of the R-symmetry group $SU(4)$ and their adjoints transform in the anti-fundamental representation of $SU(4)$. The amplitudes of this theory can be expressed in terms of three-dimensional supertwistor variables. Here we give a short description of the spinor-helicity formalism in three dimensions, for detailed discussion see \cite{Bargheer:2009qu, Bargheer:2010hn, Huang:2010rn}.

In three dimensions an on-shell null momentum can be expressed in bi-spinor notation as:
\eq
p_i^{\alpha\beta}=p_i^\mu(\sigma_\mu)^{\alpha\beta}=\lambda_i^\alpha\lambda_i^\beta
\eqe
where $\alpha,\beta$ are the indices of spinors transforming as doublets under $SO(2,1)=SL(2,R)$, and $i$ labels the external legs. This gives three components due to the symmetrization of the spinor indices. The relationship between the spinor inner products and momentum inner products is
 \eq
 \langle ij\rangle=\epsilon_{\alpha\beta}\lambda^\alpha_i\lambda^\beta_j,\;\;\langle ij\rangle^2=-2p_i\cdot p_j.
\eqe
Since $\mathcal{N}$=6 is not maximal, the on-shell multiplet is contained in two superfields:
\eqa
\nonumber\Phi(\eta)&=&\phi^4+\eta^A\psi_A+\frac{1}{2}\epsilon_{ABC}\eta^A\eta^B\phi^C+\frac{1}{3!}\epsilon_{ABC}\eta^A\eta^B\eta^C\psi_4\\
\nonumber\Psi(\eta)&=&\bar{\psi}^4+\eta^A\bar{\phi}_A+\frac{1}{2}\epsilon_{ABC}\eta^A\eta^B\bar{\psi}^C+\frac{1}{3!}\epsilon_{ABC}\eta^A\eta^B\eta^C\bar{\phi}_4.
\eqae
Although the R-symmetry group is SO(6)=SU(4), only the U(3) subgroup is manifest since the Grassmann variables $\eta^A$ have U(3) indices, i.e. $A=1,2,3$. The variables $\lambda$ and $\eta$ can be viewed as half of the supertwistor in three dimensions, which transforms in the fundamental representation of OSp(6$|$4). We will refer to the space parameterized by $(\lambda_i,\eta_i)$ as the on-shell space.

Note that the three-point interactions of this theory contain gauge fields. Since these are Chern-Simons gauge fields, they contain no dynamical degrees of freedom. Hence, only amplitudes with an even number of legs are non-trivial on-shell.

The four-point superamplitude reads:
 \eq
 A^{ABJM}_4(1,2,3,4)=\frac{\delta^3(P)\delta^6(Q)}{\langle2,1\rangle\langle1,4\rangle}=-\frac{\delta^3(P)\delta^6(Q)}{\langle2,3\rangle\langle3,4\rangle},
 \eqe
 where
 \eq
 \delta^3(P)=\delta^3(\sum^4_ip_i),\;\delta^6(Q)=\prod^3_{A=1}\delta(\sum^4_i \lambda^\alpha_i\eta^A_i)\delta(\sum^4_i \lambda_{i\alpha}\eta^A_i).
 \eqe
 At four-point, the spinor inner products have the following relationships:
\eq
\frac{\langle12\rangle}{\langle34\rangle}=\frac{\langle23\rangle}{\langle14\rangle}=\frac{\langle13\rangle}{\langle42\rangle}=\pm1.
\eqe
Using these relationships, the four-point amplitude can be written in a form similar to the one of $\mathcal{N}$=4 super Yang-Mills:
 \eq
 A^{ABJM}_4(1,2,3,4)=\frac{\delta^3(P)\delta^6(Q)}{\sqrt{\langle12\rangle\langle23\rangle\langle34\rangle\langle41\rangle}}.
 \eqe

The six-point amplitude is an object with Grassmann degree nine and has been shown, along with the four-point amplitude, to possess Yangian symmetry~\cite{Bargheer:2010hn}.\footnote{The form of the Yangian algebra along with the ordinary superconformal generators are given in appendix \ref{yg}.} The Grassmann degree of an $n$-point amplitude can be determined by the requirement that the amplitude vanishes under the U(1) generator $r^A\,_A=\sum^n_{i=1}\eta^A_i\frac{\partial}{\partial\eta^A_i}+\frac{3}{2}$. Thus in ABJM, an $n$-point amplitude has Grassmann degree $\frac{3}{2}n$ and one finds that for there are no MHV-like amplitudes for $n>4$. In other words for $n>4$ there are no amplitudes of the form:
 \eq
 \frac{\delta^3(P)\delta^6(Q)}{\sqrt{\langle12\rangle\langle23\rangle\cdot\cdot\langle n1\rangle}}.
 \eqe
\section{The dual space}
\begin{figure}
\begin{center}
\includegraphics[scale=0.8]{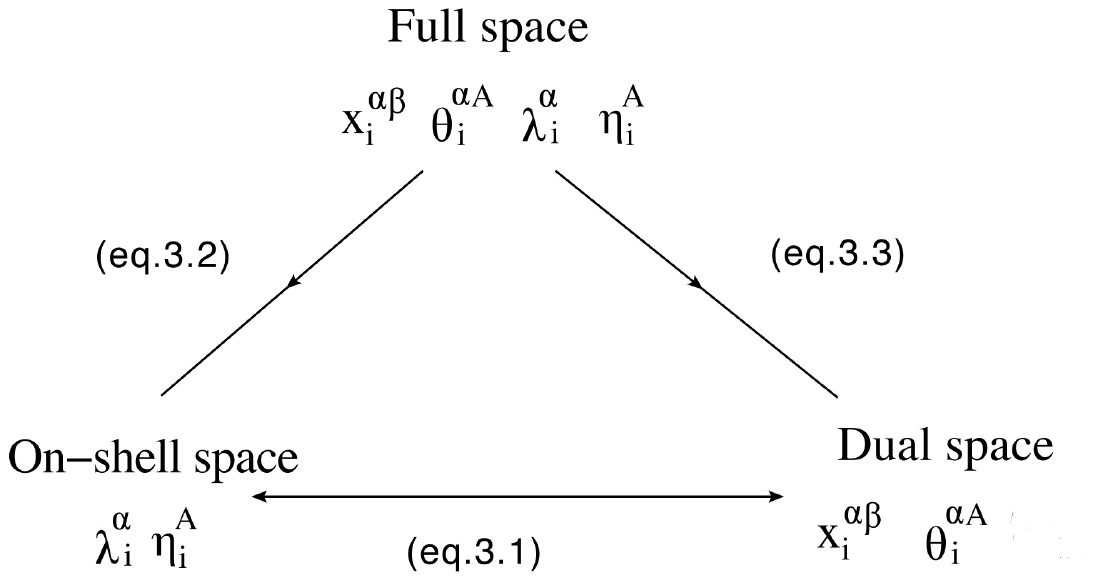}
\caption{Summary of superspace for $\mathcal{N}=6$ Chern-Simons theory. In section
6, we will show that the dual space requires three additional Grassman-even coordinates.}
\label{dualspace}
\end{center}
\end{figure}

The dual superconformal invariance of $\mathcal{N}$=4 super Yang-Mills amplitudes becomes manifest once one translates from the ``on-shell space", parameterized by $(\eta,\lambda,\tilde{\lambda})$, to the ``dual space" parameterized by $(x, \theta, \tilde{\lambda})$~\cite{restore}. Here we begin with a similar transformation to dual coordinates:
\eqa
\nonumber x_{i,i+1}^{\alpha\beta}&\equiv&x_i^{\alpha\beta}-x_{i+1}^{\alpha\beta}=p^{\alpha\beta}_i=\lambda^\alpha_i\lambda^\beta_i\\
\theta^{A\alpha}_{i,i+1}&\equiv&\theta^{A\alpha}_i-\theta^{A\alpha}_{i+1}=q_i^{\alpha A}=\lambda_i^\alpha\eta_i^A,
\label{hyper}
\eqae
where $x_{n+1}\equiv x_1,\;\theta_{n+1}\equiv \theta_1$. In these new coordinates, (super)momentum conservation is trivially satisfied. Note that $(x,\theta)$ should not be identified with the usual (super)space-time, since they would have incorrect mass dimensions. Eq.(\ref{hyper}) defines a hyperplane within the full space $(x_i,\theta_i,\lambda_i,\eta_i)$. The amplitudes have support on this hyperplane. One can translate from the dual coordinates back to the on-shell space via
\eqa
\nonumber&&x_i^{\alpha\beta}=x_1^{\alpha\beta}-\sum^{i-1}_{k=1}\lambda^\alpha_k\lambda^\beta_k,\\
&&\theta^{A\alpha}_i=\theta^{A\alpha}_1-\sum^{i-1}_{k=1}\lambda^\alpha_k\eta_k^A.
\eqae
Note that $x_1^{\alpha\beta}$ and $\theta^{A\alpha}_1$ parameterize the ambiguity that arises from the fact that eqs.(\ref{hyper}) are invariant under a constant shift in the dual coordinates. Furthermore, the hyperplane equations lead to the following relationships:
\eqa
\nonumber &&(x_{i,i+1})^{\alpha\beta}\lambda_{i\beta}=0,\quad \lambda_i^\alpha=\frac{(x_{i,i+1})^{\alpha\beta}\lambda_{i+1\beta}
}{\langle i,i+1\rangle}=\frac{(x_{i,i+1})^{\alpha\beta}\lambda_{i+1\beta}
}{\sqrt{-x^2_{i,i+2}}},\\
&&\theta^{A\alpha}_{i,i+1}\lambda_{i\alpha}=0,\quad\quad \eta_i^A=\;\frac{\theta^{A\alpha}_{i,i+1}\lambda_{i+1\alpha}}{\langle i,i+1\rangle}=\;\frac{\theta^{A\alpha}_{i,i+1}\lambda_{i+1\alpha}}{\sqrt{-x^2_{i,i+2}}}.
\label{trans}
\eqae
Given $(x_i,\theta_i)$, one can obtain all the other $\lambda$'s and $\eta$'s. In particular, after fixing $x_1$, the $\lambda$ coordinates can be determined using first relation in eq.(\ref{hyper}). After solving for the $\lambda$ coordinates, the $\eta$ coordinates can then be determined using the last relation in eq.(\ref{trans}). At this stage, the superspace for $\mathcal{N}=6$ Chern-Simons theory can be summarized in figure \ref{dualspace}. There is a similar picture for 4D $\mathcal{N}$=4 super Yang-Mills given in~\cite{restore}.

We now deduce the transformation properties of the full space under dual conformal transformations. The dual translation and part of the dual supersymmetry are trivial:
\eq
P_{\alpha\beta}=\sum^n_{i=1}\frac{\partial}{\partial x_i^{\alpha\beta}},\;Q_{\alpha A}=\sum^n_{i=1}\frac{\partial}{\partial \theta^{\alpha A}}.
\eqe
Note that these generators are consistent the hyperplane constraints in eq.(\ref{hyper}).\footnote{All dual generators are referred to using capital letters.} Since the dual conformal boost generator can be obtained from the translation generator in combination with inversion, it is sufficient to know how the various variables transform under inversion. This information can be deduced by requiring compatibility of the known transformation rules
\eq
I[x_i^{\alpha\beta}]=\frac{x_i^{\alpha\beta}}{x_i^2}=-(x_i^{-1})^{\alpha\beta},\quad I[\theta_i^{A\alpha}]=-(x_i^{-1})^{\alpha\beta}\theta^A_{i\beta},
\label{delta}
\eqe
with eq.(\ref{trans}).

For later convenience, we first we note that
\eqa
\nonumber I[(x_{i,i+1})^{\alpha\beta}]&=&-\left((x^{-1}_i)-(x^{-1}_{i+1})\right)\\
\nonumber &=&(x^{-1}_{i+1})^\beta\,_\gamma(x_{i,i+1})^{\gamma\rho}(x_i^{-1})_\rho\,^{\alpha}\\
&=&(x^{-1}_{i})^\beta\,_\gamma(x_{i,i+1})^{\gamma\rho}(x_{i+1}^{-1})_\rho\,^{\alpha}.
\eqae
From the first line of eq.(\ref{trans}), one can then deduce that
\eq
I[(x_{i,i+1})^{\alpha\beta}\lambda_{i\beta}]=0\Longrightarrow I[\lambda_{i\beta}]=\alpha_i(x_{i+1})_{\beta\gamma}\lambda_i^\gamma,
\eqe
where $\alpha_i$ is a proportionality constant. This constant can be fixed through the compatibility of the second relationship in the first line of eq.(\ref{trans}) with inversion:
\eq
I[\lambda^\alpha_i]=I\left[\frac{(x_{i,i+1})^{\alpha\beta}\lambda_{i+1\beta}}{\langle i,i+1\rangle}\right]\Longrightarrow \alpha^2_{i}=\frac{1}{(x_{i+1})^2(x_i)^2}.
\eqe
Thus we arrive at
\eqa
\nonumber I[\lambda_{i\beta}]&=&\frac{(x_{i+1})_{\beta\gamma}\lambda_i^\gamma}{\pm\sqrt{(x_{i+1})^2(x_i)^2}}=\frac{(x_{i})_{\beta\gamma}\lambda_i^\gamma}{\pm\sqrt{(x_{i+1})^2(x_i)^2}}\\
\nonumber I[\lambda^\beta_{i}]&=&\frac{(x_{i+1})^{\beta\gamma}\lambda_{i\gamma}}{\mp\sqrt{(x_{i+1})^2(x_i)^2}}=\frac{(x_{i})^{\beta\gamma}\lambda_{i\gamma}}{\mp\sqrt{(x_{i+1})^2(x_i)^2}}.
\eqae
Note that this leads to
\eq
I[\langle i,i+1\rangle]=\frac{\langle i,i+1\rangle}{\sqrt{(x_{i+2})^2(x_i)^2}}.
\label{3.9}
\eqe
Alternatively, eq.(\ref{3.9}) can be derived using the following identification in three dimensions:
\eqa
\nonumber\langle i,i+1\rangle^2&=&-2k_i\cdot k_{i+1}=-(x_{i,i+2})^2\\
\nonumber &\Longrightarrow& I[\langle i,i+1\rangle^2]=I[-(x_{i,i+2})^2]=\frac{-(x_{i,i+2})^2}{(x_{i+2})^2(x_i)^2}.\\
\eqae
Note that the spinor inner product $\langle ij\rangle$ transforms covariantly under inversion only when $i=j\pm1$. Therefore only these spinor inner products can be used to construct dual conformal objects.

Finally, for the fermionic variable $\eta$, one has:
\eq
I[\eta_i^A]=I\left[\frac{\theta^{A\alpha}_{i,i+1}\lambda_{i+1\alpha}}{\langle i,i+1\rangle}\right]\Longrightarrow I[\eta_i^A]=-\sqrt{\frac{x_i^2}{x^2_{i+1}}}\left[\eta_i^A+(x_i^{-1})^{\alpha\beta}\theta_{i\beta}^A\lambda_{i\alpha}\right].
\label{eta}
\eqe

\section{Dual conformal symmetry of ABJM amplitudes }
\subsection{Four-point amplitude}
Equipped with the transformation rules of various objects, we will now proceed to show that the four-point tree amplitude of the ABJM theory is dual conformal covariant. First we rewrite the four-point amplitude, derived in \cite{Bargheer:2010hn}, in the dual space ($x,\theta$). Note that prior to the identification of $x_{n+1}\equiv x_1,\;\theta_{n+1}\equiv \theta_1$, the hyperplane constraints imply that
$$\sum^n_{i=1}\lambda^\alpha_i\lambda^\beta_i=x^{\alpha\beta}_1-x^{\alpha\beta}_{n+1},\;\;\;\sum^n_{i=1}\lambda^\alpha_i\eta^A_i=\theta^{\alpha A}_1-\theta^{\alpha A}_{n+1}.$$
In the dual space the (super)momentum delta functions become
\eq
\delta^3(\sum_{i=1}^n p_i)=\delta^3(x_1-x_{n+1}),\quad\delta^6(\sum_{i=1}^4\eta^A_{i}\lambda^\alpha_i)=\delta^6(\theta_1-\theta_{n+1}).
\eqe
The inversion properties of the delta functions follow from the definition $\int d^3x_1\delta^3(x_1-x_5)=1$ and $I[\int d^3x_1]=\int\frac{d^3x_1}{x^6_1}$. We then have:
\eq
I[\delta^3(x_1-x_5)]=x_1^6\delta^3(x_1-x_5),\quad I[\delta^6(\theta_1-\theta_5)]=x_1^{-6}\delta^6(\theta_1-\theta_5),
\eqe
where the inversion property of the fermionic delta function is derived from eq.(\ref{delta}) on the support of $\delta^3(x_1-x_5)$. Interestingly, in three dimensions, it is only for $\mathcal{N}$=6 that the momentum and supermomentum delta functions combine to give an invariant under inversion. This is in agreement with \cite{Bargheer:2010hn}, where it was shown that Yangian invariance is only present for the $\mathcal{N}$=6 theory.\footnote{There is also Yangian symmetry in the $\mathcal{N}$=8 model~\cite{Huang:2010rn}, but it is only in a trivial sense.}

The four-point amplitude can now be written as
\eq
A^{ABJM}_{4}=\frac{\delta^3(x_1-x_5)\delta^6(\theta_1-\theta_5)}{\langle12\rangle\langle41\rangle}.
\eqe
Its transformation property under inversion is straightforward:
\eq
I[A^{ABJM}_{4}]=A^{ABJM}_4\sqrt{x^2_1x^2_2x^2_3x^2_4}.
\eqe
Note the similarity with $\mathcal{N}$=4 super Yang-Mills where one has
\eq
I[A^{\mathcal{N}=4}_{4}]=A^{\mathcal{N}=4}_4(x^2_1x^2_2x^2_3x^2_4).
\eqe
This is not surprising since we can rewrite $A_4^{ABJM}$ as
\eq
A^{ABJM}_{4}=\frac{\delta^3(x_1-x_5)\delta^6(\theta_1-\theta_5)}{\sqrt{x^2_{1,3}}\sqrt{x^2_{2,4}}},
\eqe
while
\eq
A^{\mathcal{N}=4}_{4}=\frac{\delta^4(x_1-x_5)\delta^8(\theta_1-\theta_5)}{x^2_{1,3}x^2_{2,4}}.
\eqe

From this, we see that the four-point tree amplitude is dual translation invariant and dual conformal boost covariant:
\eq
K^{\alpha\beta}A^{ABJM}_{4}=IP^{\alpha\beta}IA^{ABJM}_{4}=-\frac{1}{2}\sum^n_{i=1}x^{\alpha\beta}_iA^{ABJM}_{4}.
\eqe
We anticipate this to hold for general amplitudes. In the section \ref{dualk}, we will show that the dual conformal boost generator is equivalent to a level-one generator of the Yangian symmetry. Since both the four- and six-point amplitudes were shown to have Yangian symmetry, this implies that dual conformal invariance holds up to six-point at tree-level.
\subsection{Dual conformal invariance of general amplitudes}
The analysis of the four-point amplitude suggests the following behavior for general $n$-point amplitudes under dual inversion:
\eq
I[A_n]=\left(\sqrt{x^2_1x^2_2\cdot\cdot x_n}\right)A_n.
\label{inverse}
\eqe
Note that if one factorizes the amplitudes as follows:
\eq
A_n=\frac{\delta^3(x_1-x_n)\delta^6(\theta_1-\theta_n)}{\sqrt{\langle 12\rangle\langle 23\rangle\cdot\cdot\cdot \langle n1 \rangle}}R_n,
\label{factor}
\eqe
the pre-factor will give rise to eq.(\ref{inverse}) under inversion, so the remaining function must be invariant,
\eq
I[R_n]=R_n.
\eqe
In addition, the pre-factor is invariant under the $J^{(1)\alpha\beta}$ level-one generator of the Yangian symmetry (as pointed out in \cite{Bargheer:2010hn}). Therefore $R_n$ should be invariant under both $J^{(1)\alpha\beta}$ and the dual conformal boost generator $K^{\alpha\beta}$. Since this factorization is natural for both Yangian and dual conformal symmetry, this suggests that they are related.

So far we have the following dual conformal building blocks:
\eq
\langle i,i+1\rangle,\quad\delta^3(x_1-x_{n+1}),\quad\delta^3(x_1-x_{n+1}),\quad\delta^6(\theta_1-\theta_{n+1}).
\eqe
Recall that an $n$-point amplitude has Grassmann degree $\frac{3n}{2}$. Therefore for amplitudes higher than four-points, one needs to supplement the fermionic delta function with additional dual conformal objects which contain fermionic variables. Since $I[\lambda]\propto\lambda$, $I[x]\propto x$, and $I[\theta]\propto\theta$ but $I[\eta]\not\propto\eta$, we will use $\lambda$, $\theta$, and $x$ to construct dual conformal covariants. Noting that $\left\langle i\right|x_{i\, i+1}= \lambda_i^{\alpha} (x_{i\, i+1})_{\alpha \beta}=0$, we see that the following objects are covariant under inversion in the dual space:
\eq
\label{lam1}
\left\langle i\right|\theta_{i},\,\,\,\left\langle i-1\right|\theta_{i}{\normalcolor .}
\eqe
Moreover, they can be generalized by inserting $x_{ij}$'s as follows:
\eq
\label{lam2}
\left\langle i\right|x_{ij}x_{jk}...x_{lm}\theta_{m},\,\,\,\left\langle i-1\right|x_{ij}x_{jk}...x_{lm}\theta_{m}{\normalcolor .}
\eqe
One can also write down an inversion covariant using two $\theta$'s:
\eq
\label{theta1}
\theta_{i}\theta_{i}
\eqe
which can be generalized by inserting $x_{ij}$'s as follows:
\eq
\label{theta2}
\theta_{i}x_{ij}x_{jk}...x_{lm}\theta_{m}{\normalcolor .}
\eqe
Since $\theta$ carries an $SL(2)$ index, it is not possible to construct inversion covariants with more than two $\theta$'s except by taking products of the above covariants.

Note that all of the objects above are manifestly invariant under dual translations $\delta x_{i}=a$. We therefore have an infinite set of dual conformal covariants. We can reduce this set by demanding invariance under half of the dual supersymmetry:
\eq
\label{susy1}
\delta\theta^{A \alpha}_{i}=\epsilon^{A \alpha}{\normalcolor .}
\eqe
It is not difficult to see that this constraint eliminates the covariants in eqs. (\ref{lam1}),(\ref{theta1}), and (\ref{theta2}), but it still possible to construct objects using covariants in eq. (\ref{lam2}). For example, using the identity $x_{pq}x_{qr}+x_{pr}x_{rq}+x_{qr}^{2}=0$, one sees that there are only four dual conformal covariants with mass-dimension three that respect half of the dual supersymmetry:
\eqa
\label{fouramigos}
\left\langle p\right|\left(x_{pq}x_{qr}\theta_{r}+x_{pr}x_{rq}\theta_{q}+x_{qr}^{2}\theta_{p}\right)^A\\
\label{fouramigos2} 
\left\langle p\right|\left(x_{pq}x_{qr}\theta_{r}+x_{pr}x_{rq}\theta_{q}+x_{qr}^{2}\theta_{p+1}\right)^A\\
\label{fouramigos3}
\left\langle p-1\right|\left(x_{pq}x_{qr}\theta_{r}+x_{pr}x_{rq}\theta_{q}+x_{qr}^{2}\theta_{p}\right)^A\\
\label{fouramigos4}
\left\langle p-1\right|\left(x_{pq}x_{qr}\theta_{r}+x_{pr}x_{rq}\theta_{q}+x_{qr}^{2}\theta_{p-1}\right)^A.
\eqae
It may be possible to generalize these objects by introducing more $x$'s between the $\lambda$'s and $\theta$'s. This would require finding analogues of the identity $x_{pq}x_{qr}+x_{pr}x_{rq}+x_{qr}^{2}=0$ which are $\mathcal{O}(x^n), n>2$.

Using the covariants described above, it is straightforward to construct invariants that respect half of the dual supersymmetry. Consider the covariant in eq.(\ref{fouramigos}), for example. We will refer to this object as $\Theta^A_{pqr}$. A straightforward calculation shows that
\eq
I[\Theta^A_{pqr}]=-\frac{\Theta^A_{pqr}}{x_q^2x_r^2\sqrt{x_p^2x^2_{p+1}}}.
\eqe
To form an invariant, one needs to cancel the factors in the denominator. This can be achieved by introducing the following objects:
\eq
\langle i|x_{ij}x_{jk}|k\rangle=\langle i|x_{ij}x_{j(k+1)}|k\rangle,\;I[\langle i|x_{ij}x_{jk}|k\rangle]=\frac{\langle i|x_{ij}x_{jk}|k\rangle}{x_j^2\sqrt{x^2_{i}x^2_{i+1}x_k^2x^2_{k+1}}}.
\eqe
Now consider the six-point superamplitude, which has Grassmann degree nine. Since six of these Grassmann degrees are contained in the fermionic delta function $\delta^6(\theta_1-\theta_{n+1})$, the remaining three are then expected to have the form $\epsilon_{ABC}\Theta^A_{pqr}\Theta^B_{pqr}\Theta^C_{pqr}$, whose inversion is given by
\eq
I[\epsilon_{ABC}\Theta^A_{pqr}\Theta^B_{pqr}\Theta^C_{pqr}]=I[\delta^3(\Theta_{pqr})]=\frac{\delta^3(\Theta_{pqr})}{x_q^6x_r^6\sqrt{x_p^6x^6_{p+1}}}.
\eqe
The six-point amplitude can then be constructed from the dual conformal invariants of the form
\eqa
\nonumber&&\frac{\delta^3(\Theta_{pqr})\langle q,q+1\rangle\langle r,r+1\rangle}{\langle p|x_{pr}x_{rq}|q\rangle \langle p|x_{pq}x_{qr}|r\rangle  x^4_{qr}\langle p,p+1\rangle},\quad\frac{\delta^3(\Theta_{pqr})}{\langle p,p+1\rangle^3(x^2_{qr})^3}.
\eqae
As explained above, there are many other dual conformal building blocks one can construct, using the dual conformal covariants in eqs. (\ref{fouramigos2}-\ref{fouramigos4}), for example. It would be interesting to work out the explicit six-point amplitude in terms of dual conformal invariant objects that respect half of the supersymmetry. Instead, we proceed by showing that the dual conformal boost generator is equivalent to the level-one Yangian generator $J^{(1)\alpha\beta}$.
\section{The dual conformal boost generator and $J^{(1)\alpha\beta}$\label{dualk}}
In this section, we construct the dual conformal boost generator in the space $(x,\lambda,\theta,\eta)$ and demonstrate that it matches the Yangian level-one generator $J^{(1)\alpha\beta}$. Recall that all coordinates are independent in the full space, and the constraints in eq.(\ref{hyper}) define hyperplanes in this space. Since the amplitudes have support on these hyperplanes, the dual generators must leave the constraint equations invariant.\footnote{While the generators should preserve the plane defined by the constraints, they are not subject to the constraints, i.e. $\frac{\partial x^{\alpha\beta}}{\partial \lambda^\gamma}=0$.}

The dual translation generator takes the usual form, i.e. $\frac{\partial}{\partial x^{\alpha\beta}}$. In appendix \ref{DCB}, we obtain the dual conformal boost generator by observing how $(x,\lambda,\theta,\eta)$ transform under an inversion-translation-inversion in the dual space:
\eq
IGI, \quad G= \frac{\partial}{\partial x^{\alpha\beta}}.
\eqe
For example, in $(x,\theta)$ space, we find that the dual conformal boost generator is given by
$$
K^{\alpha\beta}=\sum^n_{i=1} x_i^{\alpha\gamma}x_i^{\beta\delta}\frac{\partial}{\partial x_i^{\gamma\delta}}+\frac{1}{2}x^{\gamma(\alpha}_{i}\theta^{A\beta)}_i\frac{\partial}{\partial \theta^{A\gamma}_i}$$
where the first term generates dual conformal boosts of $x$, the second term generates dual conformal boosts of $\theta$, and $A^{(\alpha\beta)}\equiv A^{\alpha\beta}+A^{\beta\alpha}$.  We can extend this definition to the on-shell superspace by adding terms so that it commutes with the hyperplane constraints modulo constraints \cite{Drummond:2009fd}.
Doing so gives
\eqa
\nonumber K^{\alpha\beta}&=&\sum^n_{i=1} x_i^{\alpha\gamma}x_i^{\beta\delta}\frac{\partial}{\partial x^{\gamma\delta}}+\frac{1}{2}x^{\gamma(\alpha}_{i}\theta^{A\beta)}_i\frac{\partial}{\partial \theta^{A\gamma}_i}+\frac{1}{4}\left(x^{\gamma(\alpha}_{i}\lambda^{\beta)}_i\frac{\partial}{\partial \lambda^\gamma_i}+x^{\gamma(\alpha}_{i+1}\lambda^{\beta)}_i\frac{\partial}{\partial \lambda^\gamma_i}\right)\\
&&+\frac{1}{4}\left(\theta^{B(\alpha}_{i}\lambda^{\beta)}_i\frac{\partial}{\partial \eta^B_i}+\theta^{B(\alpha}_{i+1}\lambda^{\beta)}_i\frac{\partial}{\partial \eta^B_i}\right).
\label{k}
\eqae
The normalization is due to $x^{\alpha\beta}=x^{\beta\alpha}$ and $\frac{\partial x^{\sigma\delta}}{\partial x^{\alpha\beta}}=\frac{1}{2}\delta^\sigma_{(\alpha}\delta^\delta_{\beta)}$. Indeed, one can check that eq.(\ref{k}) preserves eq.(\ref{hyper}).

As argued in sec.(4.2), the amplitudes are covariant under $K^{\alpha\beta}$:
\eq
K^{\alpha\beta}A_n=-\frac{1}{2}(\sum^n_{i=1} x^{\alpha\beta}_i)A_n.
\label{amp}
\eqe
One can derive this from our conjecture that the amplitudes transform as eq.(\ref{inverse}) under inversion. Alternatively, this follows from the factorized form given in eq.(\ref{factor}). In particular, since $R_n$ is dual conformal invariant, $K$ only acts on the string of spinor inner products. Noting that
\eqa
\nonumber&&K^{\alpha\beta}\langle ii+1\rangle=K^{\alpha\beta}\sqrt{-x^2_{i,i+2}}\\
\nonumber&=&\frac{1}{2}\left(x^{\alpha\beta}_i+x^{\alpha\beta}_{i+2}\right)\langle ii+1\rangle\\
\eqae
gives
\eq
K^{\alpha\beta}\frac{1}{\sqrt{\langle 12\rangle\langle 23\rangle\cdot\cdot\cdot \langle n1\rangle}}=-\frac{1}{2}\frac{\sum^n_{i=1}x^{\alpha\beta}_i}{\sqrt{\langle 12\rangle\langle 23\rangle\cdot\cdot\cdot \langle n1\rangle}},
\eqe
and so we arrive at eq.(\ref{amp}). Therefore, under the redefined generator
\eq
\tilde{K}^{\alpha\beta}=K^{\alpha\beta}+\frac{1}{2}(\sum^n_{i=1} x^{\alpha\beta}_i)
\eqe
we have $\tilde{K}^{\alpha\beta}A_n=0$.

Now we will demonstrate that the dual conformal boost is equivalent to a level-one Yangian generator when acting on on-shell amplitudes. Since the amplitudes can be written purely in terms of $(\lambda_i,\eta_i)$, we only consider the part of the dual conformal boost generator which acts on this space. Thus we have
\eqa
\nonumber&&\tilde{K}^{\alpha\beta}A_n=0\\
\nonumber&&\Longrightarrow  \left[\sum^n_{i=1} \frac{1}{4}\left(x^{\gamma(\alpha}_{i}\lambda^{\beta)}_i\frac{\partial}{\partial \lambda^\gamma_i}+x^{\gamma(\alpha}_{i+1}\lambda^{\beta)}_i\frac{\partial}{\partial \lambda^\gamma_i}\right)+\frac{1}{4}\left(\theta^{B(\alpha}_{i}\lambda^{\beta)}_i\frac{\partial}{\partial \eta^B_i}+\theta^{B(\alpha}_{i+1}\lambda^{\beta)}_i\frac{\partial}{\partial \eta^B_i}\right)+\frac{1}{2}x^{\alpha\beta}_i\right]A_n=0.
\eqae
Next, we trade $x$ and $\theta$ for $\lambda$ and $\eta$ using
\eq
x_i^{\alpha\beta}=x_1^{\alpha\beta}-\sum^{i-1}_{k=1}\lambda^\alpha_k\lambda^\beta_k,\quad\theta^{A\alpha}_i=\theta^{A\alpha}_1-\sum^{i-1}_{k=1}\lambda^\alpha_k\eta_k^A.
\eqe
In this way, all of the dual coordinates can be replaced except $x_1$ and $\theta_1$, however the terms containing these variables take the form
\eqa
&&\sum^n_{i=1}\frac{1}{2}x^{\gamma(\alpha}_{1}\lambda^{\beta)}_i\frac{\partial}{\partial \lambda^\gamma_i}+\frac{1}{2}\theta^{B(\alpha}_{1}\lambda^{\beta)}_i\frac{\partial}{\partial \eta^B_i}+\frac{n}{2}x^{\alpha\beta}_1\\
\nonumber&=&x^{\gamma(\alpha}_{1}\frac{1}{2}\left(m^{\beta)}\,_\gamma+\delta^{\beta)}\,_\gamma d\right)+\frac{1}{2}\theta^{B(\alpha}_{1}q^{\beta)}_B.
\eqae
Since $m,q,d$ are the usual Lorentz, supersymmetry, and dilatation generators under which the amplitudes are invariant, these terms vanish on the amplitudes. The remaining terms, which we denote as $\tilde{K}^{'\alpha\beta}$, are
\eqa
\nonumber\tilde{K}^{'\alpha\beta}&=&-\sum^n_{i=1}\left[ \frac{1}{4}\left(\sum^{i-1}_{k=1}\lambda_k^{\gamma}\lambda_{k}^{(\alpha}\lambda_i^{\beta)}\frac{\partial}{\partial \lambda^\gamma_i}+\sum^{i}_{k=1}\lambda_k^{\gamma}\lambda_{k}^{(\alpha}\lambda_i^{\beta)}\frac{\partial}{\partial \lambda^\gamma_i}\right)\right.\\
\nonumber&&\left.+\frac{1}{4}\left(\sum^{i-1}_{k=1}\lambda_k^{(\alpha}\eta^B_k\lambda^{\beta)}_i\frac{\partial}{\partial \eta^B_i}+\sum^{i}_{k=1}\lambda_k^{(\alpha}\eta^B_k\lambda^{\beta)}_i\frac{\partial}{\partial \eta^B_i}\right)+\frac{1}{2}\sum^{i-1}_{k=1}\lambda_k^{\alpha}\lambda^{\beta}_k\right].\\
\eqae
To relate this object to a level-one generator, we will write it in terms of the ordinary superconformal generators given in appendix \ref{yg}:
\eqa
\nonumber\tilde{K}^{'\alpha\beta}&=&-\sum^n_{i=1}\left[ \frac{1}{2}\sum^{i-1}_{k=1}p_k^{\gamma(\alpha}(m_i^{\beta)}\,_{\gamma}+\delta^{\beta)}\,_{\gamma}(d_i-1/2))+\frac{1}{4}p_i^{\gamma(\alpha}(m_i^{\beta)}\,_{\gamma}+\delta^{\beta)}\,_{\gamma}(d_i-1/2))\right.\\
\nonumber&&\left.+\frac{1}{2}\sum^{i-1}_{k=1}q_k^{B(\alpha }q^{\beta)}_{iB}+\frac{1}{4}q_i^{B(\alpha }q^{\beta)}_{iB}+\frac{1}{2}\sum^{i-1}_{k=1}p_k^{\alpha\beta}\right]\\
\nonumber&=&-\frac{1}{2}\sum^n_{k<i}\left[ p_k^{\gamma(\alpha}(m_i^{\beta)}\,_{\gamma}+\delta^{\beta)}\,_{\gamma}d_i)+q_k^{B(\alpha }q^{\beta)}_{iB}\right]\\
\nonumber&&-\frac{1}{4}\sum_{i=1}^n \left[p_i^{\gamma(\alpha}(m_i^{\beta)}\,_{\gamma}+\delta^{\beta)}\,_{\gamma}d_i)+q_i^{B(\alpha }q^{\beta)}_{iB}\right]+\frac{1}{4}p^{\alpha\beta}.
\eqae
At this point, it's convenient to add the following term (which vanishes on amplitudes):
\eq
\Delta\tilde{K}^{'\alpha\beta}=\frac{1}{4}\sum_{k=1}^n\left[p_k^{\gamma(\alpha}(m^{\beta)}\,_{\gamma}+\delta^{\beta)}\,_{\gamma}d)+q_k^{B(\alpha }q^{\beta)}_{B}\right]-\frac{1}{4}p^{\alpha\beta}.
\eqe
We finally arrive at
\eqa
&&\tilde{K}^{'\alpha\beta}+\Delta\tilde{K}^{'\alpha\beta}=-\frac{1}{4}\sum^n_{k<i}\left[ p_k^{\gamma(\alpha}(m_i^{\beta)}\,_{\gamma}+\delta^{\beta)}\,_{\gamma}d_i)+q_k^{B(\alpha }q^{\beta)}_{iB}-(i\leftrightarrow k)\right]\\
\nonumber&=&-\frac{1}{4}\sum^n_{k<i}\left[ (m_i^{(\alpha}\,_{\gamma}+\delta^{(\alpha}\,_{\gamma}d_i)p_k^{\gamma\beta)}-q_i^{B(\alpha }q^{\beta)}_{kB}-(i\leftrightarrow k)\right],
\eqae
which is indeed the level-one generator $J^{(1)\alpha\beta}$ given in \cite{Bargheer:2010hn}.
\section{Dual superconformal generators and new dual coordinates}
\subsection{The new dual coordinates $y^{AB}$}
Now that we have established dual conformal symmetry, we would like to extend this to dual OSp(6$|$4) superconformal symmetry by constructing $\mathcal{N}=6$ dual supersymmetry generators. If we follow what was done for $\mathcal{N}=4$ super Yang-Mills, however, we immediately encounter a difficulty: half of the supercharges are inconsistent with the constraints in eq.(\ref{hyper}). By analogy with $\mathcal{N}$=4 super Yang-Mills, the dual supersymmetry generators should be defined as
\eqa
\nonumber Q_{A\alpha}&=&\sum^n_{i=1}\frac{\partial}{\partial \theta_i^{A\alpha}},\\
\nonumber Q^A_{\alpha}&=&\sum^n_{i=1}\theta_i^{A\beta}\frac{\partial}{\partial x_i^{\alpha\beta}}+\frac{1}{2}\eta_i^A\frac{\partial}{\partial\lambda_i^\alpha}.\\
\label{firstdef}
\eqae
The first supersymmetry charge preserves both conditions in eq.(\ref{hyper}), and generates the transformation in eq.(\ref{susy1}). On the other hand, while the second charge preserves the $x$-space constraint, it violates the $\theta$-space constraint:
\eq
 Q^A_{\alpha} (\theta_i-\theta_{i+1})^{B\beta}=0,\quad Q^A_{\alpha}\lambda_i^\beta\eta_i^B=\frac{1}{2}\delta_\alpha^\beta\eta_i^A\eta_i^B\neq0.
\eqe
With a little thought, one can see that there are no terms which can be added to $Q^A_{\alpha}$ to cancel this ``anomaly".

Note that the anomaly encountered above is proportional to the single site generator $r_i^{AB}=\eta^A_i\eta^B_i$. When summed over all $i$, this gives a generator of the SU(4) R-symmetry. This suggests that we should introduce three Grassmann-even coordinates in the dual space which correspond to the generator $r^{AB}$, just like $x^{\alpha\beta}$ corresponds to $p^{\alpha\beta}$, and $\theta^{A\alpha}$ correspond to $q^{A\alpha}$. Hence we introduce three Grassmann-even coordinates, $y^{AB}=-y^{BA},\;\;A=1,2,3$, which are related to the on-shell twistor space as follows:
\eq
y^{AB}_{i,i+1}=y_i^{AB}-y_{i+1}^{AB}=\eta_i^A\eta_i^B.
\label{seconddef}
\eqe
Note that these coordinate satisfy the following pseudo light-like condition:
\eq
y^{AB}_{i,i+1}y^{CD}_{i,i+1}=\eta_i^A\eta_i^B\eta_i^C\eta_i^D=0.
\eqe
We call it pseudo since there are no invariant tensors to contract the indices to form a scalar.

Before we demonstrate that these coordinates enable us to construct the remaining dual superconformal generators, we give two arguments for their existence and their dependence on the $\eta^A$s. We note that their existence has already been suggested in~\cite{Bargheer:2010hn}.
\begin{itemize}
  \item If we didn't have the coordinates $\theta_i^{A\alpha}$, $Q^\alpha_A$ would only contain the second term in eq.(\ref{firstdef}) and would therefore violate the $x$-space constraint:
  \eq
  Q^{\alpha A}(x_i-x_{i+1})^{\beta\gamma}=0,\quad Q^{\alpha A}\lambda_i^\beta\lambda_i^\gamma=\frac{1}{2}\delta_\alpha^{(\beta}\eta_i^A\lambda_i^{\gamma)}\neq0.
  \eqe
    Note that the ``anomaly" in this case is the site generator for supersymmetry, $q_i=\eta_i^A\lambda_i^\alpha$. The resolution is to introduce a set of new coordinates $\theta^{A\alpha}_i$ and a new constraint equation for these coordinates, notably the second constraint in eq.(\ref{hyper}). With these new coordinates, the charge $Q^{A\alpha}$ can then be modified to take the form in eq.(\ref{firstdef}) so that the $x$-space constraint will be preserved. In principle, introducing new constraints may generate new ``anomalies" which can only be removed by introducing another set of new coordinates. The hope is that this process will terminate at some point. Luckily it does.
  \item Another hint which motivates introducing the $y$ coordinates is provided by the structure of a fermionic level-one generator derived in \cite{Bargheer:2010hn}:
\eq
J^{(1)\alpha A}=\sum_{i<j}\left(q_i^{\beta A}(m_j^{\alpha}\,_{\beta}+\delta^{\alpha}\,_{\beta}d_j)+q_i^{\alpha}\,_B r_j^{BA}-q_i^{\alpha B}r_j^A\,_B-s_{i\beta}^Ap_j^{\beta\alpha}-(i\leftrightarrow j)\right)
\label{fermilevel1}
\eqe
When written in on-shell superspace, the term $q_i^{\alpha}\,_B r_j^{BA}$ takes the form
\eq
q_i^{\alpha}\,_B r_j^{BA}=\lambda_i^\alpha\frac{\partial}{\partial \eta^B_i}\eta_j^B\eta_j^A.
\eqe
The only way this term can correspond to a dual superconformal generator is if one introduces the $y$ parameter. In particular, it should correspond to a generator of the form
\eq
y_i^{AB}\lambda^\alpha_i\frac{\partial}{\partial \eta^B_i}.
\eqe

\end{itemize}

After introducing the new coordinates, we alter the second supercharge as follows:
\eq
Q^{*A}_{\alpha}=\sum^n_{i=1}\theta_i^{A\beta}\frac{\partial}{\partial x_i^{\alpha\beta}}+\frac{1}{2}\eta_i^A\frac{\partial}{\partial\lambda_i^\alpha}+\frac{1}{2}y_i^{AB}\frac{\partial}{\partial \theta^{B\alpha}_i}.
\eqe
Now it is straightforward to see that the hyperplane constraint is preserved \footnote{Note that the full space is now $(x,\theta,y,\lambda,\eta)$. Although eq.(\ref{seconddef}) defines a hyperplane, all variables are taken to be independent in the full space. In particular, $\frac{\partial \eta_i^C}{\partial y_i^{AB}}=0$.}
 \eq
 Q^{*A}_{\alpha} (\theta_i-\theta_{i+1})^{B\beta}=Q^{*A}_{\alpha}\lambda_i^\beta\eta_i^B=\frac{1}{2}\delta_\alpha^\beta\eta_i^A\eta_i^B.
\eqe
Furthermore, the $y$-space constraint is also preserved and so no other additional coordinates are needed. Note that the $y$-space constraint must be respected by all generators. This implies the following deformation of the dual conformal boost generator:
\eq
K^{*\alpha\beta}=\tilde{K}^{\alpha\beta}+\frac{1}{2}\sum_{i=1}^{n}\theta_i^{A(\alpha}\theta_i^{B\beta)}\frac{\partial}{\partial y_i^{AB}}.
\label{point}
\eqe
Since the new terms do not act on the on-shell space, the dual conformal boost generator is still equivalent to a level-one Yangian generator when acting on on-shell amplitudes.

The appearance of R-coordinates like $y^{AB}$ in the superconformal generators is natural for non-chiral superspaces. For example, in four-dimensional $\mathcal{N}=2$ harmonic superspace~\cite{Galperin:1985zv}, the conformal boost generator also has a term containing R-coordinates, and this term has the same form as the additional term in eq.(\ref{point}).\footnote{We thank E. Sokatchev for pointing this out.} Since chiral superspace does not exist in three dimensions, one should expect R-coordinates to play some role in the superconformal algebra.

So far we've been able to construct the dual generators $K^{*\alpha\beta},Q^{*A}_{\alpha},Q_{\alpha A}, P^{\alpha\beta}$ (which all leave the hyperplane constraints invariant). This is sufficient to generate the entire OSp(6$|$4) dual superconformal algebra. Note that when acting on amplitudes written in terms of the on-shell space, the generators $Q_{\alpha A}$ and $P^{\alpha\beta}$ trivially vanish while $Q^{*A}_{\alpha}$ is equivalent to the original special supersymmetry generator $s^A_{\alpha}$. Hence only the vanishing of $K^{*\alpha\beta}$ on the amplitudes provides a new constraint, which we've shown is satisfied because $K^{*\alpha\beta}$ is equivalent to the level-one generator $J^{(1)\alpha\beta}$.

It is interesting to see which of the remaining dual superconformal generators imply new constraints, and if the ``non-trivial" generators are all equivalent to level-one generators. Since $Q^{*A}_{\alpha}$ is equivalent to $s^A_{\alpha}$ when restricted to the on-shell space, one can deduce that
\eqa
\nonumber &&[K^{*\beta\gamma},Q^{*A}_{\alpha}]\left|_{os}=[J^{(1)\beta\gamma},s^A_{\alpha}]=\delta^{(\beta}_{\alpha}S^{*\gamma)A}\left|_{os}\right.\right.\\
&\Longrightarrow& S^{*\alpha A}\left|_{os}=J^{(1)\alpha A}\right.,
\eqae
where $|_{os}$ means the generator is restricted to on-shell space and $S^{*\alpha A}=S^{\alpha A}+\sum^n_{i=1}\theta_i^{\alpha A}$. Thus one concludes that $S^{*\alpha A}$ implies a new constraint and matches the  level-one generator $J^{(1)\alpha A}$ when acting on on-shell amplitudes. Using similar arguments, one finds that
\eqa
\nonumber1.&&S^{\alpha}_A\left|_{os}=q^{\alpha}_A\right.\\
\nonumber2.&&R^{AB}\left|_{os}=J^{(1) AB}\right.\\
\nonumber&&{\rm since}\;\;\{S^{\alpha A},Q^B_{\beta}\}\left|_{os}=[J^{(1)\alpha A},s^B_{\beta}]=\delta^\alpha_\beta R^{AB}\left|_{os}\right.\right. \\
\nonumber3.&&R_{AB}=\frac{\partial}{\partial y^{AB}}\\
\nonumber4.&&R^A\,_B\left|_{os}=r^A\,_B\right.\\
\nonumber5.&&M^\alpha\,_\beta\left|_{os}=m^\alpha\,_\beta\right.\\
\nonumber6.&&D\left|_{os}=d\right..
\eqae
Thus the non-trivial generators are $\{K^{*\alpha\beta}, S^{*A\alpha},R^{AB}\}$, the generators which act trivially on the amplitudes are $\{P_{\alpha\beta},Q_{\alpha A}, R_{AB}\}$, and the remaining generators are equivalent to the original superconformal generators when restricted to the on-shell space. In  section \ref{ok} we will explicitly match $S^{\alpha A}\left|_{os}\right.$ and $J^{(1)\alpha A}$ with the help of the $y^{AB}$ coordinates. There is a similar correspondence between the level-one generator $J^{(1)AB}$ and the dual R-symmetry generator $R^{AB}$, which we demonstrate in appendix \ref{R}.
\subsection{Geometric interpretation of new coordinates}\label{geom}

The coordinates $y^{AB}$ can be viewed as parameterizing the half-coset ${\rm SU(4)}/{\rm U(3)}_+$. Half-cosets are constructed as follows~\cite{Hatsuda:2002wf}: first one takes a group $G$ (here SU(4)) and mods out a certain subgroup $G_0$ (here the U(3)). Next one selects a U(1) generator from $G_0$. The remaining generators in the coset can then be divided according to their charge with respect to the chosen U(1). In particular, the positively charged generators are denoted $G_+$, and the negatively charged generators are denoted $G_-$. In our case, the three $G_+$ generators actually form an abelian subalgebra.

Since the R-symmetry of the field theory corresponds to the isometries of $CP^3$, eq (\ref{seconddef}) suggests that the $y^{AB}$ coordinates should be associated with three commuting Killing vectors in $CP^3$. In Appendix \ref{cpkill}, we compute the Killing vectors of $CP^3$ and match them with the R-symmetry generators which act on on-shell amplitudes (provided in Appendix \ref{R}). The fact that type IIB string theory on $AdS_5 \times S^5$ is self-dual after T-dualizing the directions corresponding to the dual $(x,\theta)$ coordinates of $\mathcal{N}=4$ super Yang-Mills suggests that type IIA string theory on $AdS_4 \times CP^3$ should be self-dual if one performs T-dualities along the translational directions of $AdS_4$ as well as three-directions in $CP^3$ (note that a similar observation was made in \cite{Bargheer:2010hn}). As we show in Appendix \ref{cpkill}, however, the Killing vectors of $CP^3$ which correspond to the R-symmetry generators $R^{AB}$ are complex. As a result, it is not clear how to implement T-duality along these directions.

\subsection{Dual conformal supersymmetry and level-one generators\label{ok}}
The dual conformal supersymmetry generators can once again be constructed by requiring consistency with all the hyperplane constraints, or by combining inversion with the dual supersymmetry. In the end we obtain
\eqa
\nonumber S^{\alpha}_A&=&\sum_i^n x_i^{\alpha\beta}\frac{\partial}{\partial \theta_i^{A\beta}} +2\theta^{\alpha B}_i\frac{\partial}{\partial y_i^{BA}}+\lambda_i^\alpha\frac{\partial}{\partial \eta_i^A},\\
\nonumber S^{\alpha A}&=&\sum_i^n \theta_i^{\alpha B}\theta_i^{\beta A}\frac{\partial}{\partial \theta^{\beta B}_i}-x_i^{\alpha\beta}y_i^{AB}\frac{\partial}{\partial \theta_i^{\beta B}}-2x_i^{\alpha \beta}\theta_i^{\gamma A}\frac{\partial}{\partial x^{\gamma \beta}_i}-2y_i^{AB}\theta_i^{\alpha C}\frac{\partial}{\partial y^{BC}_i}\\
\nonumber&&+\frac{1}{2}\left[(\theta_i^{\alpha B}+\theta_{i+1}^{\alpha B})\eta_i^A\frac{\partial}{\partial \eta^B_i}-(\theta_i^{\beta A}+\theta_{i+1}^{\beta A})\lambda_i^\alpha\frac{\partial}{\partial \lambda^\beta_i}-(x^{\alpha\beta}_i+x^{\alpha\beta}_{i+1})\eta^A_i\frac{\partial}{\partial \lambda_i^\beta}\right.\\
\nonumber&&\left.-(y_i+y_{i+1})^{AB}\lambda_i^\alpha\frac{\partial}{\partial \eta_i^B}\right].\\
\eqae
Note that $S^{\alpha}_A$ has a simpler form than $S^{\alpha A}$ because it comes from commuting the dual conformal boost generator with $Q_{\alpha A}$, which is simpler than $Q^{\alpha A}$.

Since the amplitudes are covariant with respect to $K^{\alpha\beta}$ (as shown in eq.(\ref{amp})), we anticipate that they also satisfy
\eq
 S^{\alpha A} A_n=-\sum^n_{i=1}\theta_i^{\alpha A}A_n.
\eqe
We will show this is true by demonstrating that when one is restricted to the on-shell space
$$S^{*\alpha A}=S^{\alpha A}+\sum^n_{i=1}\theta_i^{\alpha A}=J^{(1)\alpha A}$$
i.e. the dual superconformal generator is equivalent to the fermionic level-one generator given in eq.(\ref{fermilevel1}). Since the four and six-point tree-level amplitudes satisfy Yangian symmetry, it follows that they also have dual superconformal symmetry.

When restricted to on-shell space, $S^{\alpha}_A$ is trivial while $S^{*\alpha A}$ becomes
\eqa
S^{*\alpha A}&=&\sum^n_{i=1}\frac{1}{2}\left[(\theta_i^{\alpha B}+\theta_{i+1}^{\alpha B})\eta_i^A\frac{\partial}{\partial \eta^B_i}-(\theta_i^{\beta A}+\theta_{i+1}^{\beta A})\lambda_i^\alpha\frac{\partial}{\partial \lambda^\beta_i}-(x^{\alpha\beta}_i+x^{\alpha\beta}_{i+1})\eta^A_i\frac{\partial}{\partial \lambda_i^\beta}\right.\\
\nonumber&&\left.-(y_i+y_{i+1})^{AB}\lambda_i^\alpha\frac{\partial}{\partial \eta_i^B}\right].
\eqae
We now follow steps similar to the ones we used to show that the dual conformal boost generator is equivalent to a level-one Yangian generator when acting on on-shell amplitudes. First we translate the dual coordinates $x_i^{\alpha\beta}$, $\theta^{A\alpha}_i$, and $y_i^{AB}$ back to the on-shell space using eq.(\ref{hyper}) and eq.(\ref{seconddef}). Once again, all of the dual coordinates can be removed except for the ones with $i=1$. These terms are
\eq
\sum^n_{i=1}\theta_1^{\alpha B}\eta_i^A\frac{\partial}{\partial \eta^B_i}-\theta_1^{\beta A}\lambda_i^\alpha\frac{\partial}{\partial \lambda^\beta_i}-x^{\alpha\beta}_1\eta^A_i\frac{\partial}{\partial \lambda_i^\beta}-y_1^{AB}\lambda_i^\alpha\frac{\partial}{\partial \eta_i^B}+\theta^{A\alpha}_1.
\eqe
Since $m_i^\alpha\,_\beta+\delta^\alpha_\beta d=\lambda_i^\alpha\frac{\partial}{\partial \lambda^\beta_i}+\delta^\alpha_\beta\frac{1}{2}$ and $r_i^A\,_B=\eta_i^A\frac{\partial}{\partial \eta_i^B}-\frac{1}{2}\delta^A_B$, we see that the above can be rewritten as
\eq
\theta_1^{\beta B}\left[-\delta^A_B (m^\alpha\,_\beta+\delta^\alpha_\beta d)+ \delta^\alpha_\beta r^A_B\right]-x^{\alpha\beta}_1q^A_\beta-y_1^{AB}q^\alpha_B,
\eqe
which vanishes on the amplitudes due to superconformal invariance. After doing so, we are left with
\eqa
\nonumber&&\sum^n_i\left\{-\left(\sum^{i-1}_{k=1}q_k^{\beta B}\right)\left[-\delta^A_B (m_i^\alpha\,_\beta+\delta^\alpha_\beta d_i)+ \delta^\alpha_\beta r_i^A\,_B\right]+\left(\sum^{i-1}_{k=1}p_k^{\alpha\beta }\right)s^A_{i\beta}+\left(\sum^{i-1}_{k=1}r_k^{AB}\right)q^{\alpha}_{iB}\right\}\\
&&-\sum^n_i\frac{1}{2}\left\{q_i^{\beta B}\left[-\delta^A_B (m_i^\alpha\,_\beta+\delta^\alpha_\beta d_i)+ \delta^\alpha_\beta r_i^A\,_B\right]-p_i^{\alpha\beta }s^A_{i\beta}+r_i^{AB}q^{\alpha}_{iB}\right\}.
\eqae
If we add the following term which vanishes on the amplitudes
\eq
\Delta S^{*\alpha A}=\frac{1}{2}\left\{q^{\beta B}\left[-\delta^A_B (m^\alpha\,_\beta+\delta^\alpha_\beta d)+ \delta^\alpha_\beta r^A\,_B\right]-p^{\alpha\beta}s^A_{\beta}+r^{AB}q^{\alpha}_{B}\right\}
\eqe
then we obtain
\eqa
\nonumber S^{*\alpha A}+\Delta S^{*\alpha A}&=&\sum_{1\leq k<i\leq n}\left\{q_k^{\beta A} (m_i^\alpha\,_\beta+\delta^\alpha_\beta d_i)-q_k^{\alpha B} r_i^A\,_B-s^A_{k\beta}p_i^{\alpha\beta }+q^{\alpha}_{kB}r_i^{BA}-(i\leftrightarrow k)\right\}
\eqae
This is equal to $-J^{(1)\alpha A}$. Hence, the fermionic level-one generator is equivalent to a dual conformal supersymmetry generator defined in the enlarged dual space $(x,\theta,y)$.

\section{Implications for loop-level amplitudes}
Up to now, our focus has been on tree-level amplitudes. While the number of explicit examples of tree-level amplitudes is small, results for loop-level amplitudes are even more limited. With the recent construction of tree-level four and six-point super amplitudes, however, it may now be feasible to study the structure of loop amplitudes utilizing generalized unitarity methods~\cite{Bern:1994zx}. Here we will simply assume that dual conformal invariance holds at loop-level, meaning that loop-level amplitudes can be written in terms of integrals whose representation in dual coordinates is conformally invariant. Note that this discussion is prior to using any regularization scheme, after which conformal symmetry is broken. Hence the integrals are really ``pseudo" conformal integrals. Again, since we do not have explicit computations of ABJM loop amplitudes beyond the one-loop correction to the four-point amplitude (which vanishes), this section is purely conjectural.

For $\mathcal{N}$=4 super Yang-Mills, it was observed in \cite{Drummond:2006rz, Bern:2006ew, Bern:2007ct, Drummond:2007aua} that the integrals which contribute to off-shell loop-level amplitudes are mostly ``pseudo" conformal integrals when translated to dual position space. Off-shell means that the momenta of the external lines are massive, i.e. $k_i^2\neq0$. This allows one to avoid infrared singularities while staying in $D=4$, and thus makes the discussion of dual conformal invariance sensible. Assuming this is also true for ABJM, we investigate which integrals should contribute under the constraint of dual conformal invariance.

\begin{figure}
\begin{center}
\includegraphics[scale=0.9]{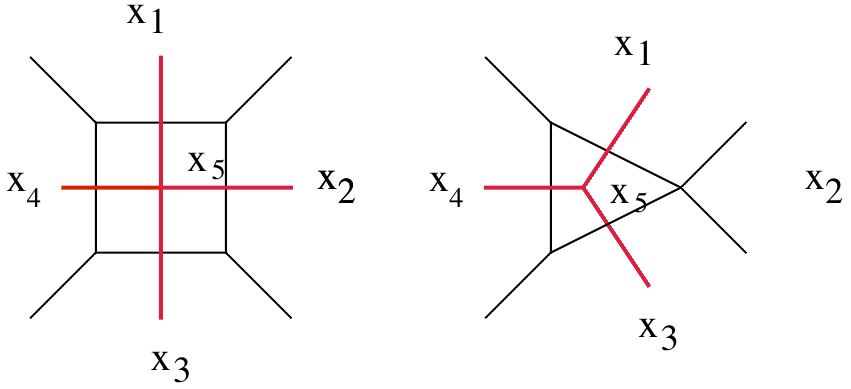}
\caption{Possible one-loop four-point integrals. }
\label{oneloop}
\end{center}
\end{figure}

Let's first consider the one-loop four-point case. As an illustration, we discuss ``pseudo" dual conformal invariance of the one-loop box diagram in fig.(\ref{oneloop}).
When written in the dual $x$ space, this amplitude takes the form:
\eq
\int d^3x_5\frac{1}{x^2_{51}x^2_{52}x^2_{53}x^2_{54}}
\eqe
Under inversion we have
\eq
I[d^Dx_5]=\frac{d^Dx_5}{(x^2_5)^D},\quad I[x^2_{ij}]=\frac{x^2_{ij}}{x^2_i x_j^2}.
\eqe
Since the integral is manifestly translationally invariant, we only need to verify that it is invariant under inversion in order to establish dual conformal invariance. This will be true if the inversion ``weight" for each coordinate $x_i$ sums to zero. For the box diagram in fig.(\ref{oneloop}), the inversion weight for $x_5$ is only zero if $D=4$. For D=3, we need three propagators to cancel the weight of the integration measure, so our only remaining option is the triangle diagram in fig.(\ref{oneloop}). Unfortunately, the weight coming from the external vertices $x_1,x_3,x_4$ cannot be canceled.

From this analysis, we can deduce rules for constructing dual conformal integrals. Given a loop diagram, we add points corresponding to positions in dual space, solid reds lines corresponding to propagators in the dual-space integral, and dashed blue lines corresponding to numerators in the dual-space integral. If a solid red line connects dual positions $i$ and $j$, this corresponds to a factor of $\frac{1}{x^2_{ij}}$ in the dual-space integral, while a blue dashed line represents a factor of $x^2_{ij}$. We then integrate over the dual coordinates that correspond to loop momenta (which happen to be $x_5$ in our example). The integral has dual conformal invariance if its diagram has the following properties:
\begin{itemize}
  \item There are three more red lines than blue lines attached to each loop momentum coordinate.
  \item There are an equal number of red and blue lines attached to each external coordinate.
\end{itemize}
In our example, the external coordinates are ($x_1,\,x_2,\,x_3,\,x_4$). From these rules, we see that it is not possible to write down a dual conformal integral for the one-loop four-point amplitude in three dimensions. This is consistent with the observation that the one-loop amplitude in mass-deformed 3D Chern-Simons theories is zero~\cite{Bargheer:2009qu}, and is simply a consequence of parity invariance. Under a parity transformation, the Chern-Simons level $k$ goes to $-k$ \cite{Bandres:2008vf}. Hence all odd-loop corrections should vanish.

At two loops, there are several integrals one can write down. We list them below:
\begin{itemize}
  \item A $$\includegraphics[scale=0.8]{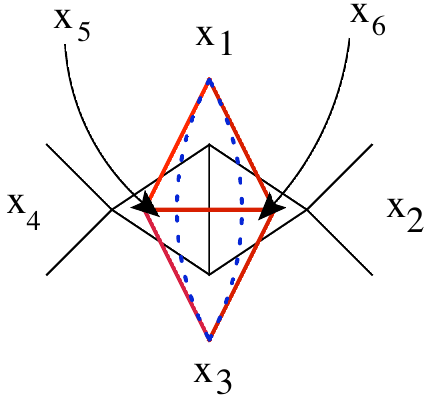}$$
  \eq
  \int d^3x_5d^3x_6\frac{x^4_{13}}{x^2_{51}x^2_{53}x^2_{56}x^2_{61}x^2_{63}}
  \eqe
  \item B $$\includegraphics[scale=0.8]{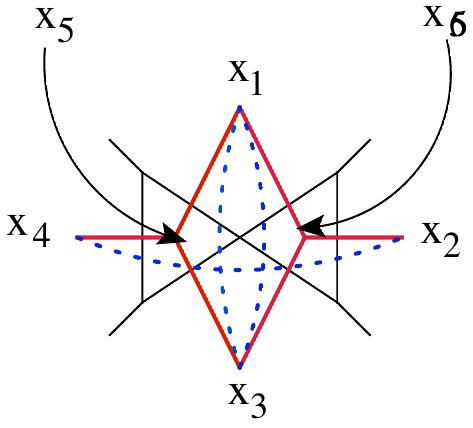}$$
  \eq
  \int d^3x_5d^3x_6\frac{x^4_{13}x^2_{42}}{x^2_{51}x^2_{53}x^2_{54}x^2_{61}x^2_{63}x^2_{62}}
  \eqe

  \item C $$\includegraphics[scale=0.8]{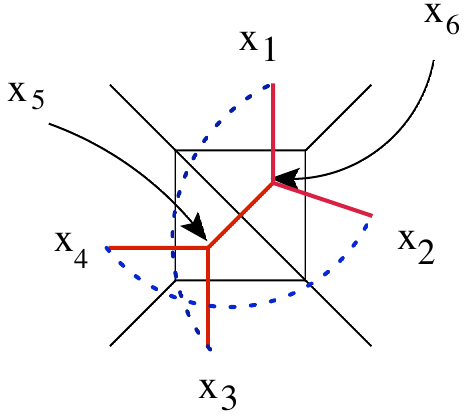}$$
  \eq
  \int d^3x_5d^3x_6\frac{x^2_{13}x^2_{42}}{x^2_{56}x^2_{53}x^2_{54}x^2_{61}x^2_{62}}
  \eqe


  \item D $$\includegraphics[scale=0.8]{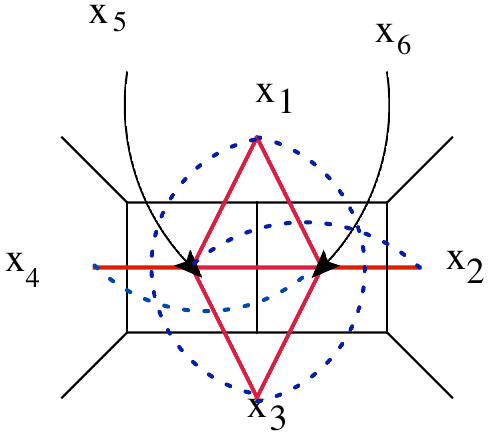}$$
  \eq
  \int d^3x_5d^3x_6\frac{x^4_{13}x^2_{52}x^2_{64}}{x^2_{51}x^2_{56}x^2_{53}x^2_{54}x^2_{61}x^2_{62}x^2_{63}}
  \eqe

\end{itemize}

Let's look at diagram A and consider it's divergence structure. In order to extract the leading divergence, we let $x_5$ and $x_6$ approach $x_1$ and write
$x_5=x_1+\rho_1$ and $x_6=x_1+\rho_2$. In this limit, the integral reduces to
$$\int \frac{\rho_1^2d\rho_1\rho_2^2d\rho_2}{\rho_1^2\rho_2^2(\rho_1-\rho_2)^2},\;\rho_1,\rho_2\rightarrow0$$
which is logarithmically divergent. Similarly, for diagrams B-D one finds
\eqa
\nonumber\;B &:&\int \frac{\rho_1^2d\rho_1\rho_2^2d\rho_2}{\rho_1^2\rho_2^2},\;(x_5,x_6\rightarrow x_1)\\
\nonumber\;C &:&\int \frac{\rho_1^2d\rho_1\rho_2^2d\rho_2}{\rho_1^2\rho_2^2},\;(x_5\rightarrow x_4,x_6\rightarrow x_1)\\
\nonumber\;D &:&\int \frac{\rho_1^2d\rho_1\rho_2^2d\rho_2}{\rho_1^2\rho_2^2(\rho_1-\rho_2)^2},\;(x_5,x_6\rightarrow x_1).\\
\eqae
Thus only the integrals $B$ and $C$ are finite. In $\mathcal{N}$=4 super Yang-Mills, it was noted that among the ``pseudo" conformal integrals, only those which are finite off-shell contribute~\cite{Drummond:2007aua}. Note that since we are off-shell, divergences cannot be attributed to infrared singularities.

In principle, there is nothing wrong with divergent off-shell integrals, since in the end one studies dimensionally regulated on-shell amplitudes. For example, after going back on-shell and regularizing, integral $A$ actually has the same structure as the four-cusp light-like Wilson-loop of pure Chern-Simons theory computed in~\cite{Henn}.\footnote{We thank Ilmo Sung for pointing this out.} Assuming that dual conformal symmetry is a well defined property for ABJM theory at loop-level, the dual conformal invariance should be well-defined off-shell, i.e. there should be no off-shell divergences. Hence, only diagrams B and C should contribute to the two-loop four-point amplitude.

Let's consider the two finite integrals more closely. They have the following form:
\eqa
\nonumber\;B &:&\int d^3pd^3q \frac{ts^2}{p^2(p+k_4)^2(p+k_3+k_4)^2q^2(q-k_1)^2(q-k_1-k_2)^2}, \quad\left(\int \frac{r^5dr}{r^6}\right)\\
\nonumber\;C &:&\int d^3pd^3q \frac{ts}{p^2(p-k_3)^2(p+q)^2q^2(q-k_1)^2},\quad \left(\int \frac{r^5dr}{r^8}\right)\\
\eqae
where the integral in the parenthesis indicates the on-shell infrared divergence. Since Diagram C has worse-than-logarithmic divergence, it requires careful treatment in dimensional regularization. Another possibility is to introduce a Yang-Mills term as a regulator, $\int-\frac{F^{\mu\nu}F_{\mu\nu}}{2g^2}$, since the dimensionful constant $g$ will serve as a cut off~\cite{Chen:1992ee}. Note that this alters the form of the gauge boson propagator. This is reminiscent of the computation of the two-loop correction to the four-cusp light-like Wilson loop in ABJM~\cite{Henn}, where the one-loop correction to the gauge boson propagator causes it to take on the form of an ordinary Yang-Mills propagator. For this reason, the two-loop correction to four-cusp Wilson loop of ABJM has the same functional form as the one-loop correction to the four-cusp Wilson loop of $\mathcal{N}=4$ super Yang-Mills.\footnote{We thank Tristan McLoughlin for discussion on this point.}

\section{Conclusions}
In this paper, we demonstrate that the Yangian invariance of the four and six-point tree-level amplitudes of ABJM implies a hidden $OSp(6|4)$ dual superconformal symmetry for these amplitudes. In order to establish this symmetry, we had to augment the dual space by three additional Grassmann-even coordinates, $y^{AB}$, which parameterize the half-coset SU(4)/U(3)$_+$. Since the generators from this half-coset form an abelian subalgebra, this corresponds to three abelian isometries of $CP^3$. The need for three additional dual coordinates was first suggested in \cite{Bargheer:2010hn}.

One way to motivate the introduction of additional coordinates is to note that in $\mathcal{N}=4$ super Yang-Mills, the dual coordinates can be matched with the dual superconformal generators which are not trivially related to the ordinary superconformal generators, notably the dual conformal boost and the dual special supersymmetry generators:
\eq
 \mathcal{N}=4\;\;{\rm sYM}:x^{\alpha\dot{\alpha}}\rightarrow K^{\alpha\dot{\alpha}},\;\theta^{\alpha I}\rightarrow S^{\alpha I}.
 \eqe
In three dimensions, this logic implies that one should also have dual coordinates corresponding to the dual R-symmetry generators $R^{AB}$ since these generators are not trivially related to the superconformal generators:
\eq
{\rm ABJM}:x^{\alpha\beta}\rightarrow K^{\alpha\beta},\;\theta^{\alpha A}\rightarrow S^{\alpha A},\;\;y^{AB}\rightarrow R^{AB}.
\eqe

Another way to motivate the need for new dual coordinates is to note that in $\mathcal{N}=4$ super Yang-Mills, the dual space is essentially the $\mathcal{N}=4$ on-shell chiral superspace. On the other hand, there is no chirality in three dimensions, so the only way to construct an on-shell space is to truncate with respect to the R-indices. Any covariant truncation would then require the introduction of some auxiliary variables parameterizing some subgroup of the R-symmetry group. While the introduction of R coordinates is usually related to the construction of off-shell superspace, they appear inside delta functions in on-shell amplitudes. For example in the $\mathcal{N}=4$ projective superspace, the four-point amplitude implicitly contains a delta function on the auxiliary coordinates~\cite{Hatsuda:2008pm}.

Note that type IIB string theory on $AdS_5 \times S^5$ is self-dual if one performs T-dualities along the directions corresponding to the dual $x$ and $\theta$ coordinates of $\mathcal{N}=4$ super Yang-Mills. This suggests that type IIA string theory on $AdS_4 \times CP^3$ might be self-dual if one performs bosonic T-dualities along three directions in $CP^3$ in addition to the three translational directions of $AdS_4$. A similar suggestion was also made in~\cite{Bargheer:2010hn}. Various groups have found that it is not possible to T-dualize the fermionic sector of the theory if one only T-dualizes the translational directions of $AdS_4$ ~\cite{Adam:2009kt,Grassi:2009yj}. In retrospect, it is difficult to see how the string theory background could be self-dual if one only T-dualizes three bosonic directions since an odd number of bosonic T-dualities would map a IIA background into a IIB background. From this point of view, it seems more natural to include three T-dualities in $CP^3$ since this would give a total of six bosonic T-dualities. By matching the $SU(4)$ R-symmetry generators with the Killing vectors of $CP^3$, we find that the Killing vectors corresponding to the dual coordinates $y^{AB}$ are complex, so it is not clear how to T-dualize these directions. It would be very interesting to determine how to implement T-duality in $CP^3$ and ultimately how to define fermionic T-duality.

Another interesting question is whether Yangian and dual superconformal symmetry hold beyond six-points at tree-level. One way to approach this issue is to construct a recursion equation and show that it preserves these symmetries. The most efficient recursion formula for theories in D$\geq$4 is the BCFW recursion relation~\cite{Britto:2004ap}. It would also be very desirable to explicitly construct the two-loop four-point amplitude of the ABJM theory using unitarity methods~\cite{Bern:1994zx} in order to see if ``pseudo" dual conformal invariance is respected, and if there is a Wilson-loop/amplitude duality for this theory.

During the completion of this paper, a manifestly superconformal invariant form of the amplitudes was proposed in~\cite{Lee:2010du} and checked against the known four-point result. It was also shown to be Yangian invariant. It would be very interesting to see how dual superconformal invariance is encoded in this formula.
\section{Acknowledgements}
The work of YH is supported by the US DOE grant DE-FG03-91ER40662 and the work of AEL is supported in part by the US DOE grant DE-FG02-92ER40701. We would like to thank Zvi Bern, Emery Sokatchev, Ilmo Sung, and Tristan McLoughlin for many suggestions, especially regarding dual special supersymmetry and loop results. We would also like to thank Till Bargheer, Johannes Henn, Harold Ita, Sangmin Lee, F. Loebbert, Juan Maldacena, Carlo Meneghelli, John H. Schwarz, and Linus Wulff for useful comments. YH would like to thank Mark Wise for the invitation as visiting scholar at Caltech.
\appendix
\section{Conventions}
We follow the conventions used in \cite{Bargheer:2010hn}. The SL(2,R) metric is
\eq
\epsilon_{\alpha\beta}=\left(\begin{array}{cc}0 & 1 \\-1 & 0\end{array}\right),\;\epsilon^{\alpha\beta}=\left(\begin{array}{cc}0 & -1 \\1 & 0\end{array}\right).
\eqe
The spinor contraction is implemented as:
\eq
\psi^\alpha \chi_\alpha =-\psi_\alpha \chi^\alpha, \quad \epsilon_{\beta\alpha}A^\alpha=A_\beta,\quad \epsilon^{\alpha\beta}A_{\beta}=A^{\alpha},\quad \epsilon^{\alpha\beta}\epsilon_{\beta\gamma}=\delta^\alpha_\gamma.
\eqe
One translates to the usual vector notation using the three-dimensional gamma matrices
\eq
x^{\alpha\beta}=x^\mu (\sigma_\mu)^{\alpha\beta},\;x^\mu=-\frac{1}{2}(\sigma^\mu)_{\alpha\beta}x^{\alpha\beta},
\eqe
with
\eq
\sigma^0=\left(\begin{array}{cc}-1 & 0 \\0 & -1\end{array}\right),\;\sigma^1=\left(\begin{array}{cc}-1 & 0 \\0 & 1\end{array}\right),\;\sigma^2=\left(\begin{array}{cc}0 & 1 \\1 & 0\end{array}\right),
\eqe
One then has,
\eq
(\sigma^\mu)_{\alpha\beta}(\sigma^\nu)^{\alpha\beta}=-2\eta^{\mu\nu},\;(\sigma^\mu)_{\alpha\beta}(\sigma_\mu)_{\gamma\delta}=\epsilon_{\alpha\gamma}\epsilon_{\beta\delta}+\epsilon_{\beta\gamma}\epsilon_{\alpha\delta}.
\eqe
We list some useful identities:
\eqa
&& A^{[\alpha\beta]}=A^{\alpha\beta}-A^{\beta\alpha}=-\epsilon^{\alpha\beta}A^\gamma\,_\gamma,\\
&& A_{[\alpha\beta]}=A_{\alpha\beta}-A_{\beta\alpha}=\epsilon_{\alpha\beta}A^\gamma\,_\gamma,\\
&& x^{\alpha\beta}x_{\beta\gamma}=-x^2\delta^\alpha_\gamma,
\eqae
where $x^2$ always represents $x^\mu x_\mu$.
\section{Ordinary superconformal and Yangian symmetry}\label{yg}
The ordinary superconformal generators of OSp(6$|$4) in the on-shell space are:
$$ p^{\alpha\beta}=\lambda^\alpha\lambda^\beta$$
$$q^{A\alpha}=\lambda^\alpha\eta^A,\;\;q^\alpha_{A}=\lambda^\alpha\frac{\partial}{\partial \eta^A}$$
$$ m^\alpha\,_\beta=\lambda^\alpha\frac{\partial}{\partial \lambda^\beta}-\delta^\alpha_\beta\frac{1}{2}\lambda^\gamma\frac{\partial}{\partial \lambda^\gamma},\;\;d=\frac{1}{2}\lambda^\gamma\frac{\partial}{\partial \lambda^\gamma}+\frac{1}{2}$$
$$ r^{AB}=\eta^A\eta^B,\;\;r^A\,_B=\eta^A\frac{\partial}{\partial \eta^B}-\delta^A_B\frac{1}{2},\;\;r_{AB}=\frac{\partial}{\partial \eta^A}\frac{\partial}{\partial \eta^B}$$
$$s^{A}_\alpha=\eta^A\frac{\partial}{\partial \lambda^\alpha},\;\;s_{\alpha A}=\frac{\partial}{\partial \lambda_\alpha}\frac{\partial}{\partial \eta^A}$$
$$k_{\alpha\beta}=\frac{\partial}{\partial \lambda^\alpha}\frac{\partial}{\partial \lambda^\beta}$$
All generators are implicitly summed over all external lines, i.e. $g=\sum^n_{i=1} g_i$. The Yangian algebra is generated by a set of level-zero and level-one generators $J^{(0)a},\;J^{(1)a}$ satisfying
\eq
[J^{(0)}_a,\;J^{(0)}_b\}=f_{ab}\,^cJ^{(0)}_c,\;\;[J^{(1)}_a,\;J^{(0)}_b\}=f_{ab}\,^cJ^{(1)}_c,
\eqe
where $f_{ab}\,^c$ is the OSp(6$|$4) structure constant and the indices can be raised and lowered using the $OSp(6|4)$ metric provided in appendix F of \cite{Bargheer:2010hn}. For the ABJM theory, the level-zero generators were identified with the superconformal generators given above, while the level-one generators are given by a bi-local product of the above single-site generators $(p_i^{\alpha\beta},q^{A\alpha}_i,q^{\alpha}_{iA},\cdot\cdot\cdot)$
\eq
J^{(1)}_a=f_a\,^{bc}\sum_{1\leq i<j\leq n}J^{(0)}_{ib}J^{(0)}_{jc}
\eqe
\section{Dual conformal boost generators using inversion properties\label{DCB}}
In the text, the dual conformal boost generator is derived from the requirement that it preserves the hyperplane-constraints. Alternatively, one can derive it by inspecting how the superspace variables transform under dual inversion, as we now describe.

Let's begin by computing the dual conformal boost generator in the space $(x,\lambda)$. Recall that $x_i^{\alpha\beta},\;\lambda_i^\alpha$ transforms under inversion as
\eq
I[x_i^{\alpha\beta}]=\frac{x_i^{\alpha\beta}}{x_i^2},\;I[\lambda_i^\alpha]=\frac{x_i^{\alpha\beta}\lambda_{i\beta}}{\sqrt{x_i^2x^2_{i+1}}}.
\eqe
The action of $K_{\gamma\delta}$ is deduced from $IP_{\gamma\delta}I=\sum^n_iI\frac{\partial}{\partial x_i^{\gamma\delta}}I$:
\eqa
\nonumber K_{\gamma\delta}x_i^{\alpha\beta}&=&\sum^n_{j=1}I\frac{\partial}{\partial x_j^{\gamma\delta}}\frac{x_i^{\alpha\beta}}{x_i^2}\\
\nonumber&=&I\left[\frac{1}{2}\frac{\delta^{(\alpha}_\gamma\delta^{\beta)}_\delta}{x_i^2}+\frac{x_{i\gamma\delta}x_i^{\alpha\beta}}{x^4_i}\right]\\
\nonumber&=&\frac{1}{2}x_i^2\delta^{(\alpha}_\gamma\delta^{\beta)}_\delta+x_{i\gamma\delta}x_i^{\alpha\beta}.\\
\eqae
Using $x_{i\gamma\delta}x_i^{\alpha\beta}=\frac{1}{2}x_{i\gamma}\,^{(\alpha}x_{i\delta}\,^{\beta)}-\frac{1}{2}x_i^2\delta^{(\alpha}_\gamma\delta^{\beta)}_\delta$, we see that
\eq
K_{\gamma\delta}\sim \sum^n_{i=1}x_{i\gamma}\,^{\alpha}x_{i\delta}\,^{\beta}\frac{\partial}{\partial x_i^{\alpha\beta}}.
\eqe

We now turn to the spinors:
\eqa
\nonumber K_{\gamma\delta}\lambda_i^{\alpha}&=&I \sum^n_{j=1}\frac{\partial}{\partial x_j^{\gamma\delta}}\frac{x_i^{\alpha\beta}\lambda_{i\beta}}{\sqrt{x_i^2x^2_{i+1}}}\\
\nonumber &=& I\left[ \frac{1}{2}\delta^{(\alpha}_\gamma\delta^{\beta)}_\delta\frac{\lambda_{i\beta}}{\sqrt{x_i^2x^2_{i+1}}}+\frac{1}{2}\left(\frac{x_{i\gamma\delta}}{x_i^2}+\frac{x_{i+1\gamma\delta}}{x_{i+1}^2}\right)\frac{x_i^{\alpha\beta}\lambda_{i\beta}}{\sqrt{x_i^2x^2_{i+1}}}\right]\\
&=&  \frac{1}{2}\delta^{\alpha}_{(\gamma} x_{i\delta)\rho}\lambda_{i}^{\rho}- \frac{1}{2}\left(x_{i\gamma\delta}+x_{i+1\gamma\delta}\right)\lambda_{i}^\alpha.
\eqae
Using $x_{i\gamma\delta}\lambda_{i}^\alpha=-\frac{1}{2}\left( \delta^\alpha_{(\delta}x_{i\gamma)}\,^\sigma\lambda_{i\sigma}+x_{i(\gamma}\,^{\alpha}\lambda_{i\delta)}\right)$ and $x_{i+1\alpha\beta}\lambda^\beta_i=x_{i\alpha\beta}\lambda^\beta_i$, we see that
\eq
K_{\gamma\delta}\lambda_i^{\alpha}=\frac{1}{4}\left(x_{i(\gamma}\,^{\alpha}+x_{i+1(\gamma}\,^{\alpha}\right)\lambda_{i\delta)}.
\eqe
Hence,
\eq
K_{\gamma\delta}\sim \sum^n_{i=1}\left[x_{i\gamma}\,^{\alpha}x_{i\delta}\,^{\beta}\frac{\partial}{\partial x_i^{\alpha\beta}}+\frac{1}{4}\left(x_{i(\gamma}\,^{\alpha}+x_{i+1(\gamma}\,^{\alpha}\right)\lambda_{i\delta)}\frac{\partial}{\partial\lambda_i^\alpha}\right]
\eqe
which is the bosonic part of eq.(\ref{k}).

The analysis for the fermionic part of the dual conformal boost generator is similar. In particular, for the fermionic coordinates of the dual superspace we have
\eqa
\nonumber
K_{\gamma\delta}\theta_{i}^{A\alpha}&=&I\left[\sum_{j=1}^{n}\frac{\partial}{\partial x_{j}^{\gamma\delta}}\frac{x_{i}^{\alpha\beta}}{x_{i}^{2}}\theta_{i\beta}^{A}\right]
\\
\nonumber &=&I\left[\left(\frac{1}{2}\delta_{(\gamma}^{\alpha}\delta_{\delta)}^{\beta}\frac{1}{x_{i}^{2}}+\frac{x_{i\gamma\delta}x_{i}^{\alpha\beta}}{x_{i}^{4}}\right)\theta_{i\beta}^{A}\right]
\\
\nonumber &=&-\frac{1}{2}\delta_{(\gamma}^{\alpha}x_{i\delta)\omega}\theta_{i}^{A\omega}+x_{i\gamma\delta}\theta_{i}^{A\alpha}.
\eqae
Using $\frac{1}{2}x_{i\,\,\,(\gamma}^{\alpha}\theta_{i\delta)}^{A}=-\frac{1}{2}\delta_{(\gamma}^{\alpha}x_{i\delta)\omega}\theta_{i}^{A\omega}+x_{i\gamma\delta}\theta_{i}^{A\alpha}$
we see that
\eq
\label{ktheta}
K_{\gamma\delta}\theta_{i}^{A\alpha}=\frac{1}{2}x_{i\,\,\,(\gamma}^{\alpha}\theta_{i\delta)}^{A}.
\eqe

Finally, let's consider the action of $K_{\gamma \delta}$ on the fermionic coordinates of the on-shell superspace:
\eq
K_{\gamma\delta}\eta_{i}^{A}=-I\left[\sum_{j=1}^{n}\frac{\partial}{\partial x_{j}^{\gamma\delta}}\left(\sqrt{\frac{x_{i}^{2}}{x_{i+1}^{2}}}\left(\eta_{i}^{A}+\frac{x_{i}^{\alpha\beta}}{x_{i}^{2}}\theta_{i\beta}^{A}\lambda_{i\alpha}\right)\right)\right].
\eqe
After taking derivatives, performing the inversion, and doing some algebra, one finds that
\eq
K_{\gamma\delta}\eta_{i}^{A}=\frac{1}{2}\left(x_{i+1}-x_{i}\right)_{\gamma\delta}\eta_{i}^{A}-\frac{1}{2}\frac{x_{i\alpha\gamma}x_{i\delta\beta}\theta_{i}^{A\alpha}\lambda_{i}^{\beta}}{x_{i}^{2}}+\frac{x_{i\gamma\delta}x_{i\alpha\beta}\theta_{i}^{A\alpha}\lambda_{i}^{\alpha}}{x_{i}^{2}}.
\eqe
Noting that $\left(x_{i+1}-x_{i}\right)_{\gamma\delta}\eta_{i}^{A}=\frac{1}{2}\left(\theta_{i+1}-\theta_{i}\right)_{(\gamma}^{A}\lambda_{i\delta)}$ and $x_{i\alpha(\gamma}x_{i\delta)\beta}\theta_{i}^{\alpha}\lambda_{i}^{\beta}=2x_{i\gamma\delta}x_{i\alpha\beta}\theta_{i}^{A\alpha}\lambda_{i}^{\beta}-x_{i}^{2}\theta_{i(\gamma}^{A}\lambda_{i\delta)}$, the above expression simplifies to
\eq
\label{keta}
K_{\gamma\delta}\eta_{i}^{A}=\frac{1}{4}\left(\theta_{i}+\theta_{i+1}\right)_{(\gamma}^{A}\lambda_{i\delta)}.
\eqe
Combining eqs.(\ref{ktheta},\ref{keta}), we see that the fermionic part of the dual conformal boost generator is
\eq
K_{\gamma\delta}\sim\sum_{i=1}^{n}\left[\frac{1}{2}x_{i\,\,\,(\gamma}^{\alpha}\theta_{i\delta)}^{A}\frac{\partial}{\partial\theta_{i}^{A\alpha}}+\frac{1}{4}\left(\theta_{i}+\theta_{i+1}\right)_{(\gamma}^{A}\lambda_{i\delta)}\frac{\partial}{\partial\eta_{i}^{A}}\right]
\eqe
which matches the fermionic part of eq.(\ref{k}).

\section{Dual R-symmetry and level-one generators\label{R}}
By requiring all hyperplane equations to be conserved, one can deduce the dual R-symmetry generator $R^{AB}$:
\eqa
\nonumber R^{AB}&=&\sum^n_{i=1}y_i^{AC}y_i^{BD}\frac{\partial}{\partial y_i^{CD}}-\frac{1}{2}y_i^{C[A}\theta_i^{\gamma B]}\frac{\partial}{\partial \theta_i^{\gamma C}}-\theta_i^{\alpha [A}\theta_i^{\beta B]}\frac{\partial}{\partial x_i^{\alpha\beta}}\\
&&-\frac{1}{4}\left(\theta_i^{\gamma [A}+\theta_{i+1}^{\gamma [A}\right)\eta_i^{B]}\frac{\partial}{\partial \lambda^\gamma_i}-\frac{1}{4}\left(y_i^{C[A}+y_{i+1}^{C[A}\right)\eta_i^{B]}\frac{\partial}{\partial \eta_i^C}.
\eqae
Note that this generator can be obtained from the dual $K^{\alpha\beta}$ by exchanging
\eq
x^{\alpha\beta}\leftrightarrow y^{AB}, \;\;\eta^A\leftrightarrow \lambda^\alpha,
\eqe
and changing signs whenever one switches from symmetrization to anti-symmetrization. Following similar steps as in the main text, one arrives at the conclusion that it is the same as the level-one generator $J^{(1)AB}$:
\eq
J^{(1)AB}\sim\sum_{1\leq k<i\leq n}q_k^{\gamma [A}s_{i\gamma}\,^{B]}+r_k^{C [A}r_{iC}\,^{B]}-(i\leftrightarrow k).
\eqe
\section{Killing vectors of $CP^3$\label{cpkill}}
The Killing vectors of $CP^{n}$ are in one-to-one correspondence with the generators of $SU(n+1)$. If we parameterize $CP^{n}$ using n+1 complex embedding coordinates $z^{I}$ satisfying
\eq
z^{\dagger}\cdot z=\sum_{I=1}^{n+1}z^{I}z_{I}^{\dagger}=1{\normalcolor },
\eqe
then the Killing vectors can be derived from n(n+2) scalar functions defined on $CP^{n}$. These functions are given by
\eq
\omega_{i}=\sum_{I,J}^{n+1}\left(T_{i}\right)_{I}^{\,\,\,\, J}z^{I}z_{J}^{\dagger}
\eqe
where $T_{i}$ are the generators of $SU(n+1)$ in the fundamental representation \cite{Hoxha:2000jf}. The Killing vectors are then given by
\eq
K_{i}^{a}=J^{ab}\partial_{b}\omega_{i}{\normalcolor }
\eqe
where $J$ is the Kahler form. Note that $J=dA$ where
\begin{equation}
A=-iz^{\dagger}\cdot dz.
\label{Aeq}
\end{equation}
Furthermore, the metric is
\eq
ds^{2}=dz^{\dagger}\cdot dz-A^{2}.
\eqe

For $CP^3$, the embedding coordinates are
$$
z^{1}=\cos\xi\cos\left(\theta_{1}/2\right)\exp\left[\frac{i}{2}\left(\psi+\phi_{1}\right)\right],\,\,\, z^{2}=\cos\xi\sin\left(\theta/2\right)\exp\left[\frac{i}{2}\left(\psi-\phi_{1}\right)\right]$$
$$
z^{3}=\sin\xi\cos\left(\theta_{2}/2\right)\exp\left[-\frac{i}{2}\left(\psi-\phi_{2}\right)\right],\,\,\, z^{4}=\sin\xi\sin\left(\theta/2\right)\exp\left[\frac{i}{2}\left(\psi+\phi_{2}\right)\right]{\normalcolor .}$$
Plugging this into eq \ref{Aeq} gives
$$A=\frac{1}{2}\left(\cos\theta_{1}\cos^{2}\xi d\phi_{1}+\cos\theta_{2}\sin^{2}\xi d\phi_{2}+\cos2\xi d\psi\right){\normalcolor .}$$
For more details about the geometry of $CP^{3}$, see Appendix B of \cite{Bandres:2009kw}.

The generators of the fundamental representation of $SU(4)$ are
$$
T_{1}=\left(\begin{array}{cccc}
0 & 1 & 0 & 0\\
1 & 0 & 0 & 0\\
0 & 0 & 0 & 0\\
0 & 0 & 0 & 0\end{array}\right),\,\,\, T_{2}=\left(\begin{array}{cccc}
0 & -i & 0 & 0\\
i & 0 & 0 & 0\\
0 & 0 & 0 & 0\\
0 & 0 & 0 & 0\end{array}\right),\,\,\, T_{3}=\left(\begin{array}{cccc}
1 & 0 & 0 & 0\\
0 & -1 & 0 & 0\\
0 & 0 & 0 & 0\\
0 & 0 & 0 & 0\end{array}\right),\,\,\, T_{4}=\left(\begin{array}{cccc}
0 & 0 & 1 & 0\\
0 & 0 & 0 & 0\\
1 & 0 & 0 & 0\\
0 & 0 & 0 & 0\end{array}\right)$$
$$
T_{5}=\left(\begin{array}{cccc}
0 & 0 & -i & 0\\
0 & 0 & 0 & 0\\
i & 0 & 0 & 0\\
0 & 0 & 0 & 0\end{array}\right),\,\,\, T_{6}=\left(\begin{array}{cccc}
0 & 0 & 0 & 0\\
0 & 0 & 1 & 0\\
0 & 1 & 0 & 0\\
0 & 0 & 0 & 0\end{array}\right),\,\,\, T_{7}=\left(\begin{array}{cccc}
0 & 0 & 0 & 0\\
0 & 0 & -i & 0\\
0 & i & 0 & 0\\
0 & 0 & 0 & 0\end{array}\right),\,\,\, T_{8}=\frac{1}{\sqrt{3}}\left(\begin{array}{cccc}
1 & 0 & 0 & 0\\
0 & 1 & 0 & 0\\
0 & 0 & -2 & 0\\
0 & 0 & 0 & 0\end{array}\right)$$
$$
T_{9}=\left(\begin{array}{cccc}
0 & 0 & 0 & 1\\
0 & 0 & 0 & 0\\
0 & 0 & 0 & 0\\
1 & 0 & 0 & 0\end{array}\right),\,\,\, T_{10}=\left(\begin{array}{cccc}
0 & 0 & 0 & -i\\
0 & 0 & 0 & 0\\
0 & 0 & 0 & 0\\
i & 0 & 0 & 0\end{array}\right),\,\,\, T_{11}=\left(\begin{array}{cccc}
0 & 0 & 0 & 0\\
0 & 0 & 0 & 1\\
0 & 0 & 0 & 0\\
0 & 1 & 0 & 0\end{array}\right),\,\,\, T_{12}=\left(\begin{array}{cccc}
0 & 0 & 0 & 0\\
0 & 0 & 0 & -i\\
0 & 0 & 0 & 0\\
0 & i & 0 & 0\end{array}\right)$$
$$
T_{13}=\left(\begin{array}{cccc}
0 & 0 & 0 & 0\\
0 & 0 & 0 & 0\\
0 & 0 & 0 & 1\\
0 & 0 & 1 & 0\end{array}\right),\,\,\, T_{14}=\left(\begin{array}{cccc}
0 & 0 & 0 & 0\\
0 & 0 & 0 & 0\\
0 & 0 & 0 & -i\\
0 & 0 & i & 0\end{array}\right),\,\,\, T_{15}=\frac{1}{\sqrt{6}}\left(\begin{array}{cccc}
1 & 0 & 0 & 0\\
0 & 1 & 0 & 0\\
0 & 0 & 1 & 0\\
0 & 0 & 0 & -3\end{array}\right).$$
Note that generators $T_1$ to $T_8$ generate an $SU(3)$ subgroup of $SU(4)$. Moreover, if we include $T_{15}$ in this subalgebra, this generates a $U(3)$ subgroup. We can then construct the half-coset described in section \ref{geom} using generators $T_9$ to $T_{14}$. In particular, $K_{a}=T_{10}+iT_9$, $K_{b}=T_{12}+i T_{11}$, and $K_{c}=T_{14}+iT_{13}$, are positively charged under the U(1) charge $\sqrt{3/2} T_{15}$. Similarly, $K_{d}=T_{10}-iT_9$, $K_{e}=T_{12}-i T_{11}$, and $K_{f}=T_{14}-iT_{13}$ are negatively charged. The half-coset ${\rm SU(4)}/{\rm U(3)}_+$ is therefore generated by $K_{a}, K_{b},$ and $K_{c}$. Using the formulas given above, the Killing vectors associated with these generators are
$$K_{a}=e^{-i\phi_{2}}\left(\frac{i}{2}\csc\left(\frac{\theta_{2}}{2}\right)\partial_{\psi}-\partial_{\theta_{2}}+i\cot\theta_{2}\partial_{\phi_{2}}\right)$$
$$K_{b}=\exp\left[\frac{i}{2}\left(\phi_{1}-\phi_{2}-2\psi\right)\right]\left[-\frac{1}{2}\sin\frac{\theta_{1}}{2}\sin\frac{\theta_{2}}{2}\partial_{\xi}\right.$$
$$-\frac{i}{16}\csc\frac{\theta_{1}}{2}\csc\frac{\theta_{2}}{2}\csc\xi\sec\xi\left(\cos\left(2\xi\right)\left(\cos\theta_{1}+\cos\theta_{2}-2\right)+\cos\theta_{1}-\cos\theta_{2}\right)\partial_{\psi}$$
$$\left.+\cos\frac{\theta_{1}}{2}\sin\frac{\theta_{2}}{2}\tan\xi\partial_{\theta_{1}}-\sin\frac{\theta_{1}}{2}\cos\frac{\theta_{2}}{2}\cot\xi\partial_{\theta_{2}}+\frac{i}{2}\csc\frac{\theta_{1}}{2}\sin\frac{\theta_{2}}{2}\tan\xi\partial_{\phi_{1}}+\frac{i}{2}\sin\frac{\theta_{1}}{2}\csc\frac{\theta_{2}}{2}\cot\xi\partial_{\phi_{2}}\right]$$
$$K_{c}=\exp\left[-\frac{i}{2}\left(\phi_{1}+\phi_{2}+2\psi\right)\right]\left[-\frac{1}{2}\cos\frac{\theta_{1}}{2}\sin\frac{\theta_{2}}{2}\partial_{\xi}+\frac{i}{8}\sec\frac{\theta_{1}}{2}\csc\frac{\theta_{2}}{2}\left(\cot\xi\left(\cos\theta_{1}+1\right)+\tan\xi\left(\cos\theta_{2}-1\right)\right)\partial_{\psi}\right.$$
$$\left.-\sin\frac{\theta_{1}}{2}\sin\frac{\theta_{2}}{2}\tan\xi\partial_{\theta_{1}}-\cos\frac{\theta_{1}}{2}\cos\frac{\theta_{2}}{2}\cot\xi\partial_{\theta_{2}}-\frac{i}{2}\sec\frac{\theta_{1}}{2}\sin\frac{\theta_{2}}{2}\tan\xi\partial_{\phi_{1}}+\frac{i}{2}\cos\frac{\theta_{1}}{2}\csc\frac{\theta_{2}}{2}\cot\xi\partial_{\phi_{2}}\right].$$
Note that these vectors are complex. As a result, it is not clear how to T-dualize along these directions.


\end{document}